\documentclass[10pt,compsoc]{IEEEtran}

\makeatletter \def\mdseries@tt{m} \makeatother 

\usepackage{booktabs} 
\usepackage{graphicx}
\usepackage[singlespacing]{setspace}
\usepackage{threeparttable}
\usepackage{multirow}
\usepackage{subcaption}
\usepackage[utf8]{inputenc}
\usepackage[english]{babel}
\usepackage{amsthm}
\usepackage{changepage} 
\usepackage{mathtools}
\usepackage{textgreek}
\usepackage[compact]{titlesec}         
\usepackage{amsmath}
\usepackage{tikz}
\usetikzlibrary{shapes,positioning,arrows,calc}

\newcommand*\circled[1]{\tikz[baseline=(char.base)]{
            \node[white,shape=circle,draw,inner sep=1pt, fill=blue] (char) {#1};}}
\DeclareFontFamily{OT1}{pzc}{}
\DeclareFontShape{OT1}{pzc}{m}{it}{<-> s * [1.1] pzcmi7t}{}
\DeclareMathAlphabet{\mathpzc}{OT1}{pzc}{m}{it}
\theoremstyle{definition}
\newtheorem{definition}{Definition}
\DeclareMathAlphabet\mathbfcal{OMS}{cmsy}{b}{n}
\usepackage{centernot}
\usepackage{enumitem}
\makeatletter
\def\algbackskip{\hskip-\ALG@thistlm}

\usepackage[backgroundcolor=white,bordercolor=blue,linecolor=blue]{todonotes}
\newcommand{\TR}[2]{\textcolor{black}{#2}}

\usepackage{lscape} 
\usepackage{xcolor}
\usepackage[ruled, vlined, linesnumbered]{algorithm2e}
\SetAlFnt{\small\sffamily}

\SetCommentSty{mycommfont}

\usepackage{amsmath, listings, amsthm, amssymb, proof, xspace}

\usepackage{hyperref}
\hypersetup{colorlinks,allcolors=black}

\theoremstyle{remark}

\lstdefinelanguage{ocl1}{
  keywords={init,Message, Pre, Post, Client,Range,Invariant,Constraints},
  keywordstyle=\color{purple}\ttfamily\bfseries,
  keywords=[2]{not, and, or, state, transition, implies, component,dbg,receipt,:,|,;},
  keywordstyle=[2]\color{black}\ttfamily\bfseries,
  sensitive=false,
  comment=[l]{//},
  morecomment=[s]{/*}{*/},
  commentstyle=\color{blue}\ttfamily,
  stringstyle=\color{red}\ttfamily,
  morestring=[b]',
  morestring=[b]"
}

\lstdefinelanguage{antlr}{
  keywords={grammar, rule,ExecRule,EOF},
  keywordstyle=\color{Purple}\ttfamily\bfseries\footnotesize,
  keywords=[2]{script:, execRule:,when:, where:, statements:, scriptStatement:,interactiveStatement:,umlrtCmd:,random:,dbgCommands:,|},
  keywordstyle=[2]\color{black}\ttfamily\bfseries\footnotesize,
  otherkeywords = {-,|},
  morekeywords = [3]{-},
  morekeywords = [4]{|},
  keywordstyle = [3]{\color{blue}}\footnotesize,
  keywordstyle = [4]{\color{blue}}\footnotesize,
  identifierstyle=\color{black}\footnotesize,
  sensitive=true,
  comment=[l]{//},
  morecomment=[s]{/*}{*/},
  commentstyle=\color{green}\ttfamily,
  stringstyle=\color{blue}\ttfamily,
  morestring=[b]',
  morestring=[b]",
  alsodigit={:}
}

\lstdefinelanguage{z3}{
	sensitive=true,
	alsoletter={\-},
	comment=[l]{;},
	keywords=[1]{
apply, assert, assert-soft, check-sat, check-sat-using, compute-interpolant,
declare-const, declare-datatypes, declare-fun, declare-map, declare-rel,
declare-sort, declare-tactic, define-sort, display, echo, eval, exit,
fixedpoint-pop, fixedpoint-push, get-assertions, get-assignment, get-info, get-
interpolant, get-model, get-option, get-proof, get-unsat-core, get-user-tactics,
get-value, help, help-tactic, labels, maximize, minimize, pop, push, query,
reset, rule, set-info, set-logic, set-option, simplify
	},
	morekeywords=[2]{
check-sat-using, declare-var, declare-rel, rule, query, set-predicate-
representation, maximize, minimize, assert-soft, assert-weighted, compute-
interpolant
	},
}

\author{Mojtaba~Bagherzadeh,
        Nafiseh~Kahani,
        and~Lionel~Briand,~\IEEEmembership{Fellow,~IEEE}
\thanks{M. Bagherzadeh and N. Kahani are with School of EECS, University of Ottawa, Ottawa, ON K1N 6N5, Canada.}
\thanks{L. Briand holds shared appointments with the SnT Centre for Security, Reliability and Trust, University of Luxembourg, Luxembourg and the school of EECS, University of Ottawa, Ottawa, ON K1N 6N5, Canada.}
}

\begin{document}

\title{Reinforcement Learning for Test Case Prioritization}

\IEEEtitleabstractindextext{%
\begin{abstract}

Continuous Integration (CI) significantly reduces integration problems, speeds up development time, and shortens release time. However, it also introduces new challenges for quality assurance activities, including regression testing, which is the focus of this work. Though various approaches for test case prioritization have shown to be very promising in the context of regression testing, specific techniques must be designed to deal with the dynamic nature and timing constraints of CI.  

Recently, Reinforcement Learning (RL) has shown great potential in various challenging scenarios that require continuous adaptation, such as game playing, real-time ads bidding, and recommender systems. Inspired by this line of work and building on initial efforts in supporting test case prioritization with RL techniques, we perform here a comprehensive investigation of RL-based test case prioritization in a CI context. To this end, taking test case prioritization as a ranking problem, we model the sequential interactions between the CI environment and a test case prioritization agent as an RL problem, using three alternative ranking models. {\color{black} We then rely on carefully selected and tailored state-of-the-art RL techniques to automatically and continuously learn a test case prioritization strategy, whose objective is to be as close as possible to the optimal one.}
Our extensive experimental analysis shows that the best RL solutions provide a significant accuracy improvement over previous RL-based work, with prioritization strategies getting close to being optimal, thus paving the way for using RL to prioritize test cases in a CI context. 

\end{abstract}

\begin{IEEEkeywords}
Continuous Integration, CI, Reinforcement Learning, Test Prioritization.
\end{IEEEkeywords}}

\maketitle

\newcommand{\para}[2]{\textit{\textbf{#1}: {#2} \\}}
\newcommand{\codee}[1]{\small\textit{#1}\normalsize}
\newcommand{\code}[1]{\small\texttt{#1}\normalsize}

\section{Introduction}
\label{sec:introdution}

Following the common practice of Continuous Integration (CI), software developers integrate their work more frequently with the mainline code base, often several times a day~\cite{duvall2007continuous}. Overall, CI significantly reduces integration problems, speeds up development time, and shortens release time~\cite{humble2010continuous}. However, it also introduces new challenges regarding quality assurance activities. Regression testing is most particularly affected due to (1) a very dynamic environment resulting from frequent changes in source code and test cases, (2) timing constraints, since regression testing should be fast enough to enable the code to be built and tested frequently. 


By default, regression testing runs all previously executed test cases to gain confidence that new changes do not break existing functionality (run-them-all approach). However, depending on the size of the code base, the number of test cases can be huge and their execution often requires many servers and can take hours or even days to complete. Test case selection and prioritization techniques remedy this issue by selecting and prioritizing a subset of test cases that are (1) sufficient to test new changes while accounting for their side effects, and (2) able to detect faults as early as possible. These techniques often rely on a mixture of code coverage analysis (e.g., ~\cite{rothermel2001prioritizing}), heuristics based on test execution history (e.g., ~\cite{kim2002history}), and domain-specific heuristics and rules (e.g.,~\cite{rothermel2001prioritizing}). Further, some researchers (e.g., ~\cite{busjaeger2016learning,bertolinolearning}) have relied on machine learning (ML) techniques in order to learn, by combining all information sources, optimal selection and prioritization heuristics. This work provides a good basis on which to address the challenges of CI regression testing. However, existing approaches must still be improved to deal with the dynamic nature and timing constraints of CI. In general, applicable test case selection and prioritization techniques must be significantly faster than the run-them-all approach to be beneficial. While the same condition holds in the context of CI, such techniques should furthermore be fast enough to avoid delays in the typically quick build cycles, as this is the main justification for CI.


Any ML based solution for test case prioritization in the context of CI needs to handle large amounts of historical data (e.g., test case and code change history) and adapt continuously to changes in the system and test suites,  reflected in newly collected data. While supervised ML techniques can deal with abundant data, their continuous adaptation to new data is impractical and time-consuming.  More specifically, the majority of current ML techniques are restricted to the classical batch setting that assumes the full data set is available prior to training, do not allow incremental learning (i.e., continuous integration of new data into already constructed models) but instead regularly reconstruct new models from scratch. This is not only very time consuming but also leads to potentially outdated models. For example, MART \cite{bertolinolearning}, the ML technique reported to be the most accurate for test case prioritization, does not support incremental learning because it is an ensemble model of boosted regression trees that is designed for static data~\cite{zhang2019incremental}.



Recently, Reinforcement Learning (RL) has shown great potential in various challenging scenarios that requires continuous adaptation, such as game playing~\cite{silver2018general}, real-time ads bidding~\cite{cai2017real}, and recommender systems~\cite{zhao2018recommendations}. Inspired by this line of work and some initial and partial efforts in supporting test case prioritization with RL techniques, we perform here a \textit{comprehensive} investigation of RL-based test case prioritization. To this end, taking test case prioritization as a ranking problem, we model the sequential interactions between the CI environment and a test case prioritization agent as an RL problem, guided by three different ranking models from information retrieval~\cite{li2011learning}: pairwise, listwise, and pointwise ranking. \TR{R3.1}{We then rely on carefully selected and tailored state-of-the-art RL techniques to automatically and continuously learn a test case prioritization strategy, whose objective is to be as close as possible to the optimal one.} 
In particular, we introduce a CI environment simulator (i.e., replayer of test execution history) based on the three ranking models, which can be used to train model parameters using available and continuously incoming test execution history from previous CI cycles, in order to prioritize test cases in subsequent cycles. The training process is adaptive in the sense that the agent is provided feedback at the end of each cycle, by replaying the execution logs of test cases, to ensure that the agent policy is efficiently and continuously adapting to changes in the system and regression test suite. Existing results regarding RL, however, show that, in terms of accuracy, it does not fare nearly as well as the best supervised ML algorithms, e.g., MART \cite{bertolinolearning}. Our main objective is therefore to benefit from the practical advantages of RL while at least retaining the prioritization accuracy of the best ML techniques. 

We have conducted extensive experiments using a variety of carefully selected RL configurations (ranking model and RL algorithm) based on eight publicly available datasets, two of them containing only execution history while the remaining six are augmented with light-weight code features. We refer to the former and latter datasets as "simple" and "enriched", respectively. The results show that, for enriched datasets, the best configurations bring a significant ranking accuracy improvement compared not only with previous RL-based work, but also with MART. Further, though the accuracy is inadequate for simple datasets regardless of the employed RL technique, we reach high accuracy for all enriched datasets, leading to test case prioritization policies that are close to the pre-determined optimal policies for each dataset. Differences in training time across configurations, tough significant, are not practically relevant in our context. Such results suggest that applying RL in practice would be beneficial when relying on adequate datasets, going beyond test execution history. 


To summarize, our work makes the following contributions towards effective and scalable test case prioritization in the context of CI.
\begin{itemize} 
   
   \item  \TR{E4, R3.2}{A \textit{comprehensive} set of solutions for the modeling of test case prioritization as an RL problem, including algorithms that precisely describe how RL can be used for each of the three ranking models from information retrieval~\cite{li2011learning}, in the context of test case prioritization and CI. These algorithms are then implemented using carefully selected, state-of-the-art RL techniques. This builds on previous work that takes a partial approach regarding the modeling, training, and implementation of RL: (1) it only uses the pointwise ranking model, (2) it relies on a small subset of RL techniques that seem to deviate from the standard, state-of-the-art algorithms provided by modern libraries~\cite{stable-baselines,baselines,hoffman2020acme}. Our work is the first that recasts pairwise and listwise ranking as an RL problem for test case prioritization. Our evaluation reveals that combining previously unused, state-of-the-art RL algorithms with pairwise ranking, results in the most accurate approach for test case prioritization.}
   
    \item Extensive experiments with a comprehensive set of carefully selected, state-of-the-art RL algorithms based on the proposed ranking models. Existing work, which we compare against, only evaluates a small subset of non-standard RL implementations based on a pointwise ranking model; in contrast, our approach evaluates 21 different RL configurations. Further, as described in Section~\ref{sec:validation}, past empirical studies on this topic have a number of issues that we attempt to address to provide more realistic results. 
    
\end{itemize}

A comprehensive set of solutions for the modeling of test case prioritization as an RL problem, including algorithms based on three ranking models. These algorithms are then implemented using state-of-the-art RL techniques.  


The rest of this paper is organized as follows. In Section~\ref{sec:background},
we define the test case prioritization problem, describe a running example, and provide background information on RL. We review related work in Section~\ref{sec:related-work} and presents three approaches for modeling test case prioritization as an RL problem in Section~\ref{sec:approch}. We present our evaluation approach and results in Section~\ref{sec:validation} and conclude the paper in Section~\ref{sec:conclusion} .

.
\renewcommand\lstlistingname{$\line(1,0){240}$ Algorithm}
\lstset{
	language=ocl1,
	extendedchars=true,
	basicstyle=\footnotesize\ttfamily,
	showstringspaces=false,
	showspaces=false,
	tabsize=2,
	breaklines=true,
	showtabs=false,
	frame=bt,
	numbers=left, 
	numberstyle=\tiny,
	mathescape=true
} 

\section{Background}
\label{sec:background}
In this section, we describe the terms and notations we use to define the test prioritization problem in our work, and describe the
RL models we rely on to support test case prioritization. 

\subsection{Test Case Prioritization}
Regression testing of a new software release is an essential software quality assurance activity. However, the regression testing of a software system with a large code base often requires the execution of a large number of test cases, which is time-consuming and resource-intensive. Test case prioritization aims to find an optimal ordering for the test case executions to detect faults as early as possible. Thus, executing a small fraction of a prioritized test suite may reduce the cost and time of testing while detecting most of the faults. In this work, our focus is regression test cases prioritization in the context of Continuous Integration (CI) of a software system where at each CI cycle, the system is built and released upon successful testing, including regression testing. 


\begin{definition}\noindent\textbf{(CI and CI Cycles).}  \label{def:ci}
We capture the $CI$ history of a software system as sequence of cycles $c_{i}, 1 < i < n $, where $c_1$ and $c_n$ refer to the first and current cycles, respectively. A cycle $c$ is a tuple $\langle T,f \rangle$, where $T$ is a set of test cases, and $f$ is a logical value that indicates whether or not the cycle has failed. The number of test cases in different cycles varies. \TR{E1, R2.6}{A cycle can fail due to several reasons, including compilation errors or a test case failure. However, in this work, we are only interested in the latter and failed cycles are, in our experimental datasets, cycles with at least one failed test case.}

\end{definition}

\begin{definition}\noindent\textbf{(Test Case Feature Records).}  \label{def:testcasefeatures}
Each test case has two feature records: execution history and code base features.  
Execution history of a test case at cycle $c_i$ is defined as a tuple $<v,e,h, a>$, where $v$ shows the execution verdict of the test case at cycle $c_i$, $e$ represents the execution time of the test case at cycle $c_i$, $h$ is a sequence of verdicts that shows the test case verdicts at prior cycles, i.e., $c_j, 1<j \leq i-1 $, and $a$ represents the test age capturing when the test case was first introduced. The execution history for each test case contains a record of executions over previous cycles. The execution verdict is considered either $1$ if the test case has failed, or $0$ if it has passed. Similar to previous work \cite{spieker2017reinforcement}, we assume the execution time ($e$) of the test case to be the average of its previous execution times.

\begin{table*}[t!]
\centering
	\caption{{\color{black}Code-based features, adopted from~\cite{bertolinolearning}}} 
	\label{tab:metrics}
    \begin{tabular}{p{1.80cm}p{10.2cm}p{4cm}}
    \hline
     \small \textbf{Type} &      \small \textbf{Features}  &      \small \textbf{Description}   \\ 
    \hline

\small \textbf{Program size} &\small  AvgLine, AvgLineBlank, AvgLineCode, AvgLineComment, CountDeclFunction, CountLine, CountLineBlank, CountLineCode,
CountLineCodeDecl, CountLineCodeExe, CountLineComment,
CountSemicolon, CountStmt, CountStmtDecl, CountStmtExe,
RatioCommentToCode  & Features related to the amount of lines of code, declarations, statements, and files  \\ \hline 
\small \textbf{McCabe’s cyclomatic complexity} &\small AvgCyclomatic, AvgCyclomaticModified, AvgCyclomaticStrict,
AvgEssential, MaxCyclomatic, MaxCyclomaticModified, MaxCyclomaticStrict, MaxEssential, MaxNesting, SumCyclomatic, SumCyclomaticModified, SumCyclomaticStrict, SumEssential  & Features related to the control flow graph of functions and methods\\ \hline
\small \textbf{Object oriented metrics} &\small CountDeclClass, CountDeclClassMethod, CountDeclClassVariable, CountDeclExecutableUnit, CountDeclInstanceMethod,
CountDeclInstanceVariable, CountDeclMethod, CountDeclMethodDefault, CountDeclMethodPrivate, CountDeclMethodProtected, CountDeclMethodPublic  & Features based on object-oriented constructs  \\ \hline

    \end{tabular}

\end{table*}

The accuracy of all ML techniques largely depends on the features they use. Solely relying on execution history, thus ignoring many relevant code-based features, such as the complexity of changes or test cases, can lead to low prioritization accuracy. Thus, we adopt code-based features from the literature~\cite{bertolinolearning} that are extracted using light-weight and incremental static analysis and repository mining techniques, and are thus applicable in a CI context. \TR{R3.3}{Code-based features such as Line of Code (LoC) are relevant predictors of test case execution time and failure occurrences. For instance, if tests t1 and t2 target source files f1 and f2 respectively, and f1 has more LoC, then the execution of t1 is likely to take longer than that of t2, because t1 targets a more complex source file.  We can make a similar argument about the probability of failure.}

\TR{R2.3, R3.3}{
Table~\ref{tab:metrics} lists the code-based features  calculated for each test case, based on the source code classes that are exercised (covered) by  test execution. Assuming dependency and coverage data for source code classes is available,  Bertolino et al.~\cite{bertolinolearning} use the following four-step process to relate code-based features to test cases and create a vector for each test case in each CI build. 
}
{
\color{black}
\begin{enumerate}
    \item The changed classes in the build are identified.  
    \item All impacted files are extracted from the dependency database based on the changed classes (output of step 1).
    \item For each test case, a subset of impacted classes  (output of step 2) covered by its execution are identified.
    \item For each test case, code-based features are calculated based on covered classes (output of step 3). When a test case covers more than one file, the features are calculated based on all of the covered classes by adding their values.
\end{enumerate}

Dependency and coverage data is collected from the source code of both test cases and the system under test using static analysis techniques (i.e., more specifically using Understand~\cite{understand}). Then such data is updated for each build based on new changes. Impact analysis and incremental updates enable efficient coverage and dependency analysis (as discussed in Section~\ref{sec:datasets}). Though static analysis tends to overestimate coverage and dependencies, such information can help improve the accuracy of ML models. 

}



\end{definition}

\begin{definition}\textbf{(Test Case Prioritization).} \label{def:prioritization} 
Regardless of underlying techniques for test case prioritization, we assume this to be a ranking function that takes in input a set of test cases' features and returns an ordered sequence in which the position (index) of the test cases shows their priority for execution, i.e., the test case with the lowest index (rank) gets executed first. 

\end{definition}

\begin{definition}\textbf{(Optimal Ranking).} \label{def:optimalranking} 
Given a set of $n$ test cases ($T$), the ranking function can produce $n!$ distinct ordered sequences. We define the optimal order ($s_o$) of a set of test cases $T$ as a sequence in which: \\
\indent $\forall t_1,t_2 \in T, \\
\indent \indent idx(s_o,t_1) < idx(s_o,t_2) \iff t_1.v > t_2.v \; or \;$ \\
\indent \indent $(t_1.v = t_2.v \; and \; t_1.e <= t_2.e)$, \\
where function $idx(s,t)$ returns the index (rank) of a test case $t$ in sequence $s$.

The above condition implies that, in the optimal ranking of the test cases, (1) all failing test cases (their verdict is one) are executed before passing test cases, (2) test cases with lower execution time are executed earlier than other test cases with the same verdict. 

We use the optimal order $s_o$ as a reference ranking (ground truth) and our goal is to find a test case ranking function whose output is as close as possible to $s_o$. \TR{R3.4}{Note that we give higher importance to the verdict than the execution time. However, depending on the context, the optimal ranking can be tuned. For instance, if one knows with confidence that failures are very rare, the optimal ranking can be tuned to give higher importance to execution time rather than the verdict.}

\end{definition}

\subsection{Illustrative Example} 
\begin{figure}
    \centering
    \includegraphics[width=8cm]{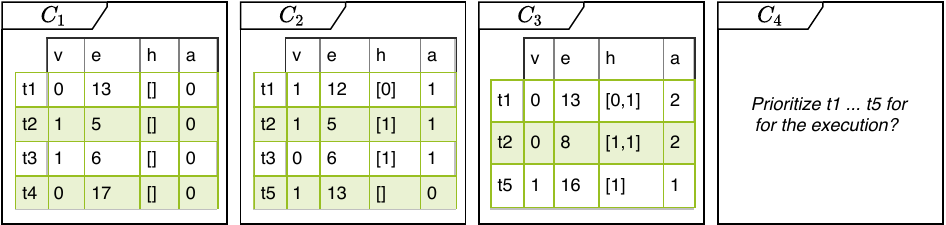}
    \caption{An artificial example of CI}
    \label{fig:ciexample}
\end{figure}
Figure~\ref{fig:ciexample} shows an artificial CI example in which only history features of test cases are included in the interest of space. The example shows three completed CI cycles, each of which contains a few test cases. As shown, (1) the number of test cases varies across cycles, (2) execution history of a test case at a specific cycle contains previous execution verdicts, and (3) the age of test cases is incremented after each cycle completion. 

An optimal ranking for a specific cycle is estimated based on test execution history. This is because test execution times and verdicts are unknown until all test case executions are completed and; therefore, the optimal ranking is unknown. 

\begin{figure}
    \centering
    \includegraphics[width=8cm]{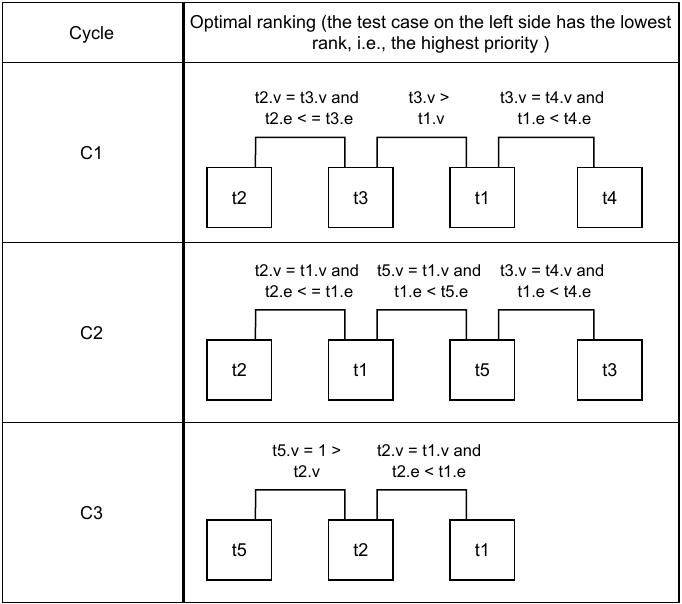}
    \caption{\color{black}Optimal rankings for the illustrative example (Fig. ~\ref{fig:ciexample})}
    \label{fig:optimalranking}
\end{figure}

\TR{E1, R1.1}{Based on our illustrative example above, we provide optimal rankings for all cycles in Figure~\ref{fig:optimalranking}.}

\subsection{Reinforcement Learning (RL)} \label{sec:reinforcement}
In RL, an agent interacts with its environment through the use of observations (states), actions, and rewards. At each interaction step $t$, the agent receives some representation of the environment’s state as input, $S_t \in S$,  where S is the set of possible states. Based on the perceived state, the agent chooses an action, $A_t \in A(S_t)$, where $A(S_t)$ is the set of actions available in state $S_t$,  to generate as output. The action selection is based on either a learned or a exploration policy. As a result, the agent receives feedback in terms of reward, which rates the performance of its previous action. 


State-of-the-art RL techniques can classified based on following properties:\\
\textbf{Model-based versus model-free.} In model-free RL algorithms, it is assumed that an agent neither has any prior knowledge of the environment (a black-box environment) nor attempts to learn the environment dynamics. In other words, the agent does not know beforehand how the environment reacts to possible actions, or what the next state and reward will be before taking an action. So, the agent needs to interact with the environment and observe its responses to devise an optimal policy for selecting an action. As mentioned earlier, the execution time and results of test case executions at a given CI cycle are unknown before their execution. Therefore, we only use model-free RL algorithms for test case prioritization. 


\textbf{Value based,  policy  based, and  actor-critic learning.}
Assuming that the Q-value is a measure of the expected reward in a state for a given action, value-based methods estimate the Q-value of possible actions for a given state and select the action with the highest value. An example of value-based methods is the Q-learning algorithm~\cite{watkins1992q} that, in its simplest form, uses a Q-table and the Bellman equation to estimate the Q-value. In its more advanced form (DQN), it uses a deep neural network to estimate the Q-value~\cite{mnih2015human}.
Policy-based methods directly search for an optimal policy. Typically, a parameterized initial policy is chosen, whose parameters are updated to maximize the expected return using either gradient-based or gradient-free optimization. An example of policy-based methods is the REINFORCE algorithm~\cite{williams1992simple}.

Each of the above methods has drawbacks and benefits. More importantly, value-based methods are often sampling efficient. But the convergence is guaranteed in very limited settings that often requires extensive hyperparameter tuning. On the contrary,  policy-based methods are stable but sample inefficient, i.e, convergence is guaranteed but at a very slow rate~\cite{nachum2017bridging,konda2000actor,precup2000eligibility}. Actor-critic methods aim at combining the strong points of actor-only (value-based) and critic-only (policy-based) methods. The critic uses an approximation architecture and simulation to learn a value function, which is then used to update the actor's policy parameters. Such methods have desirable and faster convergence properties compared to value-based and policy-search based methods~\cite{konda2000actor}. In this work, we do not exclude any algorithm based on their learning method because there is no evidence regarding the superiority of a certain method in all contexts. 

\textbf{Action and observation space.} The action space specifies how the agent can act on its environment, while the observation space specifies what the agent can know about its environment. The latter is referred to as feature space in ML. Both the observation and action spaces come in discrete and continuous forms. In the simplest form, an observation can only be a real number (e.g., the position of the agent)  but it can also be more complex and high-dimensional (e.g., RGB matrix of observed pixel values). With a discrete action space, the agent decides which distinct action to perform from a finite action set, whereas with a continuous action space, actions are predicted and expressed as a real-valued vector. While most of the RL algorithms do not impose constraints on the observation space, not all of them support both discrete and continuous action spaces, and therefore their application is restricted according to the problem's action space. Further, it is possible for the action space to be a vector of continuous or discrete values.  
As we will discuss later (Sec. \ref{sec:approch}), we use three different approaches for modeling test case prioritization, each of them having a different form of action space that limits our choice of algorithms. 


\textbf{On-policy vs off-policy.} There are two types of policy learning methods, namely on-policy (e.g., SARSA) and off-policy. In on-policy learning, agent attempts learn a policy that is close to the exploration strategy, i.e, the learned policy is influenced by the exploration strategy. While, in off-policy learning (e.g., Q-learning), the learned policy is independent of the exploration strategy, i.e., exploration during the learning phase is not based on the learned policy~\cite{sutton2018reinforcement}.




\begin{table}[t!]
\centering
	\caption{Model-free, state-of-the-art RL algorithms} 
	\vspace{-.3cm}
	\label{tab:algos}
    \begin{tabular}{|c|c|c|c|}
    \hline
     \small \textbf{Algo.} &      \small \textbf{Lear.}  &      \small \textbf{On/Off}  &      \small \textbf{Act.}  \\ 
    \hline

     \small \textbf{DQN~\cite{mnih2013playing}}   &\small   Value&  Off-policy&  Dis 
  \\
   
     \hline
   
   \small \textbf{DDPG~\cite{DDPG}}     & \small   Policy&  Off-policy &  Cont 
     \\ \hline
     
        \small \textbf{A2C~\cite{A2C}}     & \small   Actor-Critic&   On-policy &  Both  
    \\ \hline
        \small \textbf{ACER~\cite{ACER}}     & \small   Actor-Critic&   Off-policy &  Dis 
    \\ \hline
        \small \textbf{ACKTR~\cite{ACKTR}}     & \small   Actor-Critic&   On-policy &  Both 
    \\ \hline
        \small \textbf{TD3~\cite{TD3}}     & \small   Policy&   Off-policy &  Cont 
     \\ \hline
        \small \textbf{SAC~\cite{SAC}}     & \small    Actor-Critic&   Off-policy &  Cont 
    \\  \hline
        \small \textbf{PPO1~\cite{PPO}}     & \small   Actor-Critic&   On-policy &  Both  \\ \hline
         \small \textbf{PPO2~\cite{PPO}}     & \small   Actor-Critic&   On-policy &  Both
   \\  \hline
        \small \textbf{TRPO~\cite{TRPO}}     & \small  Actor-Critic&   On-policy &  Both  

   \\  \hline
     \multicolumn{4}{l}{\footnotesize \textit{Cont}: continuous, \textit{Dis}: discrete, \textit{Both}: discrete and continuous}
    \end{tabular}

\end{table}

\subsection{State-of-the-art RL Algorithms and Frameworks}
Several model-free RL algorithms have been proposed over the last few years that advance the state of the art, e.g., Deep Deterministic Policy Gradient (DDPG)~\cite{DDPG},  Deep Q-Networks (DQN), Advantage Actor-Critic (A2C)~\cite{A2C}. Further, several open-source research frameworks provide reusable implementations of the state-of-the-art algorithms, e.g., Acme~\cite{hoffman2020acme}, Stable Baselines~\cite{stable-baselines}, and OpenAI Baselines~\cite{baselines}. In this work, our focus is on the application of RL techniques rather than devising new RL techniques. Thus, we rely on the  \textit{Stable Baselines framework} and state-of-the-art algorithms which it provides. \textit{Stable Baselines} is the improved version of \textit{OpenAI Baselines}, with more comprehensive documentation and support for more algorithms compared to other frameworks. A list of the supported algorithms that match our problem as we discuss in Sections ~\ref{sec:approch} and \ref{sec:validation}, as well as their properties, are shown in Table~\ref{tab:algos}. Note that all of the above algorithm use deep neural networks (DNNs) to capture policies.

\section{Related work}
\label{sec:related-work}
Test prioritization for regression testing has
long been an active area of research \cite{yoo2012regression,suleiman2017survey,khatibsyarbini2018test}. Existing work can be categorized into two groups: heuristic-based and ML-based test prioritization. 

\textbf{Heuristic-based Test Prioritization.} The proposed methods of this group have typically used heuristics based on information such as code coverage \cite{rothermel2001prioritizing, rothermel1999test}, models \cite{tahat2012regression,korel2005test}, history \cite {kim2002history,fazlalizadeh2009incorporating, park2008historical}, and requirements \cite{srikanth2005system,srikanth2005economics}. The main drawback of these methods, especially in a CI context, is that they are not adaptive to quickly changing environments.  

A large body of existing work focuses on using code coverage information and the analysis of code modifications to order test cases. Coverage-based techniques stem from the idea that early maximization of structural coverage can increase the chances of early maximization of fault detection~\cite{kim2002history}. Some of the structural coverage measures include statement coverage \cite{rothermel1999test}, functions/methods coverage \cite{rothermel2001prioritizing}, and modified condition/decision coverage  \cite{jones2003test}. Rothermel et al. \cite{rothermel1999test} presented several approaches for prioritizing test cases and reported empirical results measuring the effectiveness of these approaches. \TR{E4, R3.4}{Overall, the coverage-based work can be grouped into two groups: total requirement coverage and additional requirement coverage~\cite{jeffrey2006test}. The former orders test cases in decreasing order of the number of statements they cover. The latter prioritize test cases in decreasing order of the number of additional statements they cover, that is statements that have not yet been covered by the previously executed test cases.}

\TR{R3.4}{Coverage information can be collected either by static or dynamic analysis. Lightweight static analysis techniques overestimate the coverage data and are not accurate~\cite{sui2020recall}. More thorough static analysis techniques (e.g., static analysis with reflection support) can significantly improve the accuracy of coverage information, but their high computation cost renders them impractical~\cite{sui2020recall}. Similarly, dynamic analysis techniques are difficult or even impossible to apply in practice, and more specifically so in a CI context. The reasons are discussed in several papers~\cite{elbaum2014techniques,spieker2017reinforcement,lima2020learning,do2020multi,memon2017taming} and summarized below:} {\color{black}
\begin{itemize}
    \item Computation Overhead: Code analysis and instrumentation take a long time to execute for a large codebase~\cite{elbaum2014techniques,do2020multi,memon2017taming}. As reported in~\cite{memon2017taming}, running a code instrumentation tool at each milestone on the codebase of Google and collecting code coverage data would impose too large an overhead to be practical.  
    \item Applicability: They are applicable only to complete sets of test cases, as they search in the space of all test cases and select/prioritize them to reach either maximum coverage or defect detection, or minimum execution cost~\cite{elbaum2014techniques}. Also, the extraction of code coverage requires traceability between code and test cases, the information that is not always available or easily accessible with system tests (i.e., black-box testing)~\cite{lima2020learning}. Further, the non-ML based techniques are often language and platform-dependent, which leads to more customization and effort. 
    \item Maintainability: Typically, high code change rates, in actively developed projects, quickly render code coverage data obsolete, requiring frequent updates~\cite{do2020multi,memon2017taming}. 
\end{itemize}

}


Several researchers \cite{korel2005test} used executable system models to select and generate test cases related to the modified parts of the system. Models are an abstraction of the actual system. Such abstractions make the model execution for the whole test suite relatively inexpensive and fast compared to the execution of the actual system \cite{korel2009experimental}.  However, the source code may change over time, resulting in the need to update the models to reflect the changes. Such  updates create overhead when relying on model-based approaches. Also, models are often extracted using source code analysis; therefore, they inherit the drawbacks of code-based approaches.
Korel et al. \cite{korel2005test} presented a model-based prioritization approach in which the original and modified system models, along with the information about the system model and its behavior, are used to prioritize test cases. 
While code-based coverage approaches are more precise compared to model-based ones, they introduce practical challenges in terms of complexity and  computational overhead to collect and analyze code coverage information.

History-based approaches rank tests based on past test execution data. These approaches are based on the idea that past test case failures are a good predictor of test cases with high probability of failure in new releases.
Kim and Porter \cite{kim2002history} proposed a history-based approach that calculates ranking scores based on the average of past execution results.
Park et al. \cite{park2008historical} proposed a history-based approach to analyze the impact of test costs and  severity of detected defects in a test suite on test prioritization.
Noor and Hemmati \cite{{noor2015similarity}} defined a class of quality metrics that estimate test case quality using their similarity to the previously
failing test cases from previous releases. Their results showed that adding similarity-based test quality metrics along with traditional test quality metrics can improve the test prioritization results.
History-based approaches are less expensive than coverage-based and  model-based approaches. However, learning optimal test case prioritization policies only based on test execution history seems difficult, specifically for complex software systems.  Also, they may not be well adapted to continuously changing testing environments with frequent changes in code and test suites. In our work, we also rely on historical information, with the differences that (1) our RL-based solution is seamlessly adaptive and can therefore deal with the dynamic nature of CI (2) we use enriched execution history with code-based features to improve the accuracy of the prioritization.

Srikanth et al. \cite{srikanth2005system} proposed a model for system-level test case prioritization from software requirement specifications. They mapped test cases to software requirements, and then prioritized the test cases based on four factors including requirements volatility, customer priority, implementation complexity, and fault-proneness of the requirements. Similar work \cite{srikanth2005economics} proposed a system-level technique for prioritization based on requirements according to four factors including customer assigned priority of requirements, developer-perceived implementation complexity, requirement volatility and fault proneness of the requirements.

Some work \cite{marijan2013test, elbaum2014techniques} proposed heuristic-based test prioritization methods tailored to CI environments. Marijan et al. \cite{marijan2013test} proposed a weighted history-based test prioritization approach called ROCKET, which orders test cases based on historical failure data, test execution time, and domain-specific heuristics. Elbaum et al. \cite{elbaum2014techniques} presented a test selection approach at the pre-submit stage  that uses time windows to track how recently test cases have been executed and revealed failures. To increase the cost-effectiveness of testing, they performed test prioritization based on the specified windows to prioritize test cases that must be executed during subsequent post-submit testing.

In contrast with these techniques, our approach uses RL to prioritize test cases. Relying on RL makes our approach seamlessly adaptive to the changing CI environment. Also, by combining various data sources (e.g., coverage, failures, execution time), we may be able to build more accurate prioritization models. 

\textbf{ML-based Test Prioritization Techniques:} Work \cite{durelli2019machine} in this category investigates the application of ML techniques to test prioritization. The motivation is to integrate data from different sources of information into accurate prediction models. Results have shown that ML techniques can provide noticeably promising results in test selection and prioritization \cite{durelli2019machine}. 

Several approaches studied the effectiveness of clustering for test prioritization. Carlson et al. \cite{carlson2011clustering} cluster test cases based on code coverage, code complexity, and fault history data. Lenz et al. \cite{lenz2013linking} grouped test cases into functional clusters derived by executing some example test cases. The test results and clusters feed ML classifiers, which produce sets of rules to classify the test cases.  The rules were used to support various tasks, including test case prioritization.

Past research  \cite{tonella2006using,lachmann2016system, busjaeger2016learning,bertolinolearning} has also proposed a number of supervised ML techniques that reduced test prioritization to a ranking problem. 
Tonella et al. \cite{tonella2006using} proposed a pairwise ranking algorithm to rank test cases based on  coverage, complexity metrics and historical data.
Busjaeger and Xie \cite{busjaeger2016learning} introduced a listwise method based on ML combined with multiple existing heuristic techniques to prioritize test cases in industrial CI environments. They used features include coverage data, test file path similarity and test content similarity, failure history, and test age.
Lachman et al. \cite{lachmann2016system} applied SVM Rank to black-box prioritization starting from test cases and failure reports in natural language (NL). 

However, supervised and unsupervised ML techniques tend to be impractical in a CI context when prediction models need to continuously and quickly adapt to new data, reflecting changes in the system and test suites. To deal with this issue, recent work has investigated the application of RL. 
In an initial attempt to apply one RL algorithm to test case prioritization in CI environments \cite{spieker2017reinforcement}, Spieker et al. prioritize test cases according to their execution time and previous execution and failure history. Their work is based on the pointwise ranking model and only uses the Q-learning RL algorithm. In contrast, in this work, we perform a comprehensive investigation of RL techniques by guiding the RL agent according to three different ranking models: pairwise, listwise, and pointwise ranking. 

In very recent work, Bertolino et al. \cite{bertolinolearning} analyze the performance of ten ML algorithms, including three RL algorithms, for test prioritization in CI. Through an experimental analysis, they show that Non-RL-based approaches to test case prioritization are more affected by code changes, while the RL-based algorithms are more robust. 
Similar to Spieker et al. \cite{spieker2017reinforcement}, Bertolino's application of RL is based on the pointwise ranking model. Their results show that their specific RL configuration is significantly less accurate compared to the best ranking algorithms based on supervised learning (e.g., MART). Further, the above RL-based work only experiment with a small subset of RL implementations that differ from state-of-the-art algorithms provided by modern libraries  ~\cite{stable-baselines,baselines}. 



This paper builds on past work by applying RL to CI regression testing. It does so by investigating all ranking models: pointwise, pairwise, and listwise. Further, for each model, we experiment with all available and applicable state-of-the-art RL algorithms, as it is difficult to a priori determine which ones will work better in a CI regression testing context. Thus, we increase our chances of obtaining accuracy results that are close to or better than the best supervised learning techniques, e.g., MART, while getting the practical benefits of RL.  

\section{Reinforcement Learning for Test Case Prioritization}
\label{sec:approch}

\textbf{An Overview.}    
We aim to develop an RL-based solution for the prioritization of test cases in the context of CI. Most existing test case prioritization solutions consider the prioritization procedure as a static process and prioritize test cases following a fixed strategy that is defined based on either heuristics or supervised ML techniques~\cite{yoo2012regression}. Here, we investigate variants of a prioritization approach capable of continuously adapting and improving its strategy as a result of its interactions with the CI environment.
We model the sequential interactions between CI and test case prioritization as an RL problem and rely on state-of-the-art RL techniques to automatically and continuously learn a test case prioritization strategy that is as close as possible to the optimal one, assuming a pre-determined optimal ranking as the ground truth. In particular, we introduce a CI environment simulator, which can be used to train the agent offline using the available test execution history before applying and updating the model online. \TR{R2.2, R2.6, R3.6}{In other words, we train an RL agent based on test execution history and code-based features from previous cycles in order to prioritize test cases in subsequent cycles. The training process is adaptive in the sense that the agent can be provided with feedback at the end of each cycle or, when the agent accuracy is below a certain threshold, execution logs of test cases can be replayed to ensure the agent policy is efficiently and continuously adapting to changes in the system and regression test suite.}

\TR{R3.2}{We rely on a typical approach for developing an RL solution in a specific context: (1) devise algorithms that precisely describe how RL can be used for each ranking model, in the context of test case prioritization and CI via replaying the test cases' execution history, (2) train an RL agent using carefully-selected, state-of-the-art RL techniques, as discussed in Section ~\ref{sec:background}. One important goal in this paper is to be as comprehensive as possible in investigating alternatives. Next, we discuss possible solutions for creating an RL environment for test case prioritization, with a focus on the formalization of action and observation spaces, reward functions, and interactions between the environment and the RL agent. We then describe the way RL techniques can be applied to train an agent in our context and discuss how the RL agent can be integrated into CI environments.} 
%






\subsection{Creation of the RL Environment}

As discussed in Section ~\ref{sec:reinforcement}, the RL agent and environment interact by passing the observation, reward, and action. The typical flow of the interaction is shown in Algorithm~\ref{code:rlflow}. First, the agent is given an initial observation by the environment. Then an episode starts, during which the agent perceives the current observation and selects an action based on the exploration strategy, which varies according to the underlying algorithm, e.g., Q-learning uses the epsilon-greedy ($\epsilon-$greedy) exploration method. An episode is a sequence of states and actions that takes an RL agent from an initial state to a final state, in which the agent task is completed. 

The selected action is passed to the environment that applies the action and returns a new observation and reward. The agent takes the reward and observation into account and updates its policy according to the underlying RL technique and most particularly the learning method (policy, value, or actor-critic based), as discussed in Section ~\ref{sec:background}. The episode ends when the task is done, regardless of success or failure. The end condition depends on the nature of the task, e.g., for an agent that plays a game, an episode ends when the game ends.

\begin{lstlisting}[numbers=left,xleftmargin=2em,frame=single,framexleftmargin=1.5em,float,mathescape=true,language=Python,caption={A Training Episode of an RL Agent},label={code:rlflow},frame=top,frame=bottom,frame=lines,morekeywords={Input,Output,bar,Function}, escapeinside={@}{@}] 
Let done be False # a flag to capture the end of episode
Let obs be a valid initial observation

while not done # an episode
  action=predict(obs) # the agent select an action
  done, reward, obs = step(action) # applying the action 
  updateAgentPolicy() # agent reinforces its policy 
    
\end{lstlisting}
To map the test case prioritization problem to RL, the details of the above-mentioned interactions (i.e., observation, action, reward, and end condition of an episode) need to be defined properly. Assuming test case prioritization to be a ranking function (Definition~\ref{def:prioritization}), the interaction details can be defined based on ranking models from the information retrieval field~\cite{li2011learning}: pointwise, pairwise, and listwise. A pointwise ranking approach takes the features of a single document and uses a prediction model to provide a relevance score for this document. The final ranking is achieved by simply sorting the documents according to these predicted scores. For pointwise approaches, the score for each document is independent from that of the other documents.

A pairwise approach orders a pair of documents at a time. Then, it uses all the ordered pairs to determine an optimal order for all documents. 
Some of the most popular Learning-to-Rank algorithms are pairwise approaches~\cite{burges2010ranknet}, e.g., RankNet, LambdaRank and LambdaMART. Listwise ranking approaches consider a complete list of documents at once and assign a rank to each document relative to other documents.  


Each ranking model has advantages and drawbacks in the context of test case prioritization, that will be further discussed below. Also, the interaction details between the \textit{Agent} and \textit{Environment} will differ based on the underlying ranking model. In the following, we discuss how each of the ranking models can be applied in the context of test case prioritization. 



\subsubsection{Listwise solution}
{
\label{sec:listwise}
Algorithm~\ref{code:listwise} shows the details of an episode of listwise ranking that starts by setting the initial observation containing a vector of all test cases' features. As discussed in Section ~\ref{sec:background}, the number of test cases varies in different cycles. However, RL, as other ML techniques, does not handle inputs of variable size. Therefore, the size of the observation space for a given CI system needs to be defined, based on the maximum number of test cases in a cycle, to allow for the trained agent to handle all cycles by using padding. For example, based on the illustrative example, the maximum number of test cases in a cycle is $4$, while the number of test cases for cycle $C_3$ is $3$. Thus, we add a dummy test case whose features are set to $-1$ into the feature record of $C_3$. We refer to the process of creating these dummy test cases as padding.

After preparing the initial observation, an episode is started, during which the agent selects the index of the test case with the highest priority (lowest rank) as an action. As shown in function $step$ of Algorithm~\ref{code:listwise}, the environment applies the action by (1) appending the selected test case to the output sequence ($s_e$), (2) updating the observation by replacing the feature record of the selected test case with the dummy test case to keep track of selected test cases, and (3) calculating a reward that is shown in function \textit{calc\_reward}.

\begin{lstlisting}[numbers=left,xleftmargin=2em,frame=single,framexleftmargin=1.5em, float,mathescape=true,language=Python,caption={RL-based Listwise Ranking},label={code:listwise},frame=top,frame=bottom,frame=lines,morekeywords={Input,Output,bar,Function}, escapeinside={@}{@}] 
Input 
Let $T$ be a @set@ of test cases 
Let $s_o$ be the optimal ranking of $T$

Output 
$s_e$ an order (a sequence) of test cases in $T$ 

Let $rank$ be 0, done be False, @and@ reward be 0 
Let $obs$ be a vectorized  $T$  # observation space
Let $action$ be an integer with @range@ [0,|$obs$|-1]
episode() # ref. Algorithm $\ref{code:rlflow}$

Function step($action$)
    $reward$ = calc_reward($action$)
    if $rank$ < |$T$|-1 and obs[$action$] is not dummy:
        append obs[$action$] to $s_e$
        $obs$[$action$] = a dummy test case # mark the test case as selected
        $rank$ = $rank$ +1
    else if $rank$ = |$T$|-1
        $done$ = True
    return $done$, $reward$, $obs$ 
        
Function calc_reward($action$)   
 if $obs[action]$ is dummy: # previously selected or padded
    $reward$ = 0
 else
    $optimal\_rank$ = idx($s_o$,$obs[action]$)
    $reward$ = 1- (norm($optimal\_rank$) - norm($action$))^2

\end{lstlisting}

\textbf{Observation and action space.} In listwise ranking, the observation space grows linearly with the number of test cases and this increases the training and prediction time. In general, dealing with large observation and action spaces is one of the main challenges with RL~\cite{dulac2019challenges,dulac2015deep}. \TR{R1.2}{In our context, in which the feature record of a test case has at least four numeric fields (Def. ~\ref{def:testcasefeatures}), for a system with 1,000 test cases the observation space grows to 4,000 numeric fields (features). This is clearly a high-dimensional observation space, with each feature having a large range of possible numeric values. Thus, an enormous amount of training data is required to ensure that there are several samples for each combination of values to train the model. Coping with this kind of growth in feature space dimensionality is an open and active research area in machine learning~\cite{donoho2000high}.
}


\textbf{Action space.}  The action space in listwise ranking is a discrete value whose range is defined by the number of test cases. That value, the action, captures the test case with the highest priority. Similar to the observation space, the action space grows linearly with the number of test cases. This can also lead to scalability issues since the agent needs to evaluate all possible actions at each step. Thus, the larger the action space, the larger the training time. In general, existing RL techniques cannot handle large discrete spaces~\cite{dulac2015deep,dulac2019challenges}. 

As an alternative, it would be possible to define the action space of the listwise ranking as a vector of either discrete or continuous values, each of which representing the rank of a test case, and  then train an agent to assign ranks of all test cases in one step. A vector of discrete or continuous values as an action space is supported by the existing RL algorithms (e.g., TRPO) and can be applied for a cycle with a small number of test cases. However, when the action space is large, training the agent (finding an optimal policy) is difficult. Some algorithms do not converge to the optimal policy, converge very slowly, or, in some cases, have
prohibitive computational requirements~\cite{weisz2018sample}. We have performed an initial experiment using a vector of discrete values as action space and tried to train an agent on a cycle with 600 test cases that took more than six hours. As a result, in our experiments, we will adopt the first solution presented above, where the action space is a single discrete value. 

\textbf{Reward function.} To calculate reward, we take the optimal test case ranking (Definition ~\ref{def:optimalranking}) as the reference and compare the assigned rank of each test case with respect to its rank in the optimal ranking. 
As shown in function \textit{calc\_reward} of Algorithm~\ref{code:listwise}, the reward function calculates the reward, as a value within $[0,1]$, for the selected action (index of a test case) based on its deviation from the optimal ranking. The agent gets the highest possible reward when the rank assigned to a test case is equal to the optimal one. The smaller the distance between the RL and optimal rankings, the higher the reward. 
Further, the agent is given the lowest reward (zero) when the agent selects a dummy test case, resulting either from the test case having been already been selected or padding.

Overall, the listwise ranking is easy to model. However, the high dimensionality of the action and observation spaces causes scalability issues. We provide more details on performance of this ranking model in Section~\ref{sec:validation}.  
}
\subsubsection{Pointwise solution.}
\label{sec:pointwise}
Algorithm~\ref{code:pointwise} shows the details of an episode of pointwise ranking that starts by converting the set of test cases to a sequence and then setting the initial observation to the first test case of the sequence. It then begins a training episode, during which the agent determines a score for the test case that is a real number between 0 and 1.  As shown in function $step$ of Algorithm~\ref{code:pointwise}, the environment applies the action by (1) saving the selected score in a temporary vector ($tmp$), (2) updating the observation by setting the next test case of the vector as observation, (3) calculating a reward that is shown in function $calc\_reward$, and (4) at the end of episodes, sort all the test cases based on their assigned scores saved in $tmp$.

\begin{lstlisting}[numbers=left,xleftmargin=2em,frame=single,framexleftmargin=1.5em,float,mathescape=true,language=Python,caption={RL-based Pointwise Ranking},label={code:pointwise},frame=top,frame=bottom,frame=lines,morekeywords={Input,Output,bar,Function}, escapeinside={@}{@}] 
Input 
Let $T$ be a @set@ of test cases 
Let $s_o$ be the optimal ranking of $T$

Output 
$s_e$ an order (a sequence) of test cases in $T$ 

Let $s_e$ be a random sequence of test cases in $T$
Let $index$ be 0, done be False, @and@ reward be 0
Let $obs$ be a vector of [$s_e$[$index$]] 
Let $tmp$ be an empty vector # keeps ranks
Let $action$ be a real number with a @range@ (0,1]
episode () ## ref. Algorithm 1

Function step(action)
    reward=calc_reward(action)
    add $action$ into $tmp$
    if $index$ < |$s_e$|-1
        $index$ = $index$+1
    else
        $done$ = True
        sort $s_e$ based on their ranks in $tmp$
    $obs$ = $s_e$[$index$]
    return $done$, $reward$, $obs$
    
Function calc_reward(action) 
    Let $optimal\_rank$ be the position of $s_e$[$index$] in optimal ranking
    $reward$ = 1- (norm($optimal\_rank$) - $action$)^2

\end{lstlisting}

 
\textbf{Observation and action space.}
An observation here is the feature record of a single test case, which is much smaller than an observation for the listwise approach. The action space is a continuous range between 0 and 1, that is the test case’s score, a real number based on which test cases are ranked. 

\textbf{Reward function.}
Similar to listwise, we take the optimal ranking as reference and calculate the reward based on the distance between the assigned and optimal ranks. However, since each test case is scored individually during training, the final rank of the test cases is not known until the end of the training episode. Thus, as shown in function \textit{calc\_reward} of Algorithm~\ref{code:pointwise}, we compute the distance by normalizing the optimal rank of test cases. Since the normalized optimal and assigned score values both range between 0 and 1, their difference provides meaningful feedback to the agent. 
An alternative is to only provide the reward at the end of the episode. However, this leads to the sparse reward issue~\cite{riedmiller2018learning} and makes the training of the agent inefficient.

Overall, the pointwise ranking is easier to model than listwise and the dimensionality of its observation space is much smaller compared to listwise.  We provide more details on the performance of this ranking model in Section ~\ref{sec:validation}.

\subsubsection{Pairwise solution.}
\label{sec:pairwise}
Algorithm~\ref{code:pairwise} shows the details of an episode of pairwise ranking that starts by setting the initial observation to a pair of test cases. It then starts a training episode, during which the agent selects either $0$ or $1$, the former denoting that the first test case in the pair has higher priority (lower rank). In general, ranking a pair is based on a comparison operator is the essential building block of the sorting algorithms, such as selection and merge sort~\cite{knuth1998art}. Thus, based on the ranking of a pair, the environment can apply any of the sorting algorithms to prioritize the test cases. For example, as shown in function $step$ in Algorithm~\ref{code:pairwise}, the environment applies the selection sort algorithm~\cite{knuth1998art} to sort all the test cases based on ranking pairs. Thus, it updates the observation based on a mechanism that the selection sort prescribes. That is, the list is divided into two parts ($idx_0$ is the splitter), the sorted part at the left of $idx_0$ (indices below $idx_0$) and the unsorted part at the right of $idx_0$. At each iteration of the sort, it finds the test cases with the highest priority and changes its position to $idx_0$ and then increases $idx_0$ and repeats this process until all test cases are sorted.
The environment also calculates the rewards as shown by \textit{calc\_reward} in Algorithm~\ref{code:pairwise}.


The required steps of each training episode are determined by the complexity order of the applied sorting algorithms. For example, the complexity order of the selection sort is quadratic in the worst case and requires $n^2$ comparisons where $n$ is the number of test cases. Therefore, each episode of the agent training requires $n^2$ steps to prioritize $n$ test cases. While this is manageable for cycles with a small number of test cases, it causes very long-running episodes for cycles with a large number of test cases, e.g., for $10,000$ test cases, an episode requires around 50 million steps to be completed. To alleviate this problem, we can adopt sorting algorithms with a lower complexity such as merge sort~\cite{knuth1998art}, the complexity of which is linearithmic ($n \; log(n) $). Thus, we also defined  another version of the pairwise environment based on the merge sort algorithm, which is similar to Algorithm~\ref{code:pairwise}. Merge sort is actually used in our experiments but it is quite long to describe ($>$ 60 lines of code) and is not necessary to understand the fundamentals of our approach. The interested reader can refer to the source code\footnote{\url{https://github.com/moji1/tp\_rl/blob/master/testCase\_prioritization/PairWiseEnv.py}}.

\begin{lstlisting}[numbers=left,xleftmargin=2em,frame=single,framexleftmargin=1.5em, float,mathescape=true,language=Python,caption={RL-based Pairwise Ranking},label={code:pairwise},frame=top,frame=bottom,frame=lines,morekeywords={Input,Output,bar,Function}, escapeinside={@}{@}] 
Input 
Let $T$ be a @set@ of test cases 
Let $s_o$ be the optimal ranking of $T$

Output 
$s_e$ an order (a sequence) of test cases in $T$ 

Set sequence $s_e$ to a be random order of test cases in $T$
Let $idx_0$ be 0, $idx_1$ be 1, done be False, @and@ reward be 0
Let $obs$ be a vector [$s_e$[$idx_0$] , $s_e$[$idx_1$]] # a pair
episode()

Function step(action)
    reward = calc_reward(action)
    # the following lines performs the selection sort in a stepped way 
    if action == 1:
        swap($s_e$, $idx_0$, $idx_1$)
    if $idx_1$ < |$s_e$|-1:
        $idx_1$ = $idx_1$ + 1
    else if  ($idx_1$ == |$s_e$| - 1) and ($idx_0$ < |$s_e$| - 2):
        $idx_0$ = $idx_0$ + 1
        $idx_1$ = $idx_0$ + 1
    else
        done = True
    $obs$ = vector of [$s_e$[$index_0$] , $s_e$[$index_1$]] 
    return done, reward, obs
    
    
Function calc_reward(action)
    Let $sel\_test\_case$ be $obs[action]$ and $nonsel\_test\_case$ be  $obs[1- action]$
    if $sel\_test\_case.v$ > $nonsel\_test\_case.v$:
        reward = 1
    else if   $sel\_test\_case.v$ < $nonsel\_test\_case.v$:
        reward = 0
    else if $sel\_test\_case$.e <= $nonsel\_test\_case.e$:
        reward = 0.5
    else 
        reward = 0
        


\end{lstlisting}


\textbf{Observation and action space.}
An observation (state) is a pair of test case feature records and is therefore much smaller than for listwise, but twice as large as pointwise. The action space is simply the set {0,1}, which is much simpler than the action spaces of both listwise and pointwise. 

\textbf{Reward function.}
If the agent gives higher priority to the only failed test case in the pair, it receives the highest reward ($1$). Otherwise, it receives no reward ($0$). Also, if both test cases in a pair have the same verdicts, then the agent receives $0.5$ as reward when it gives higher priority to the test case with less execution time. Otherwise, the agent receives no reward ($0$). Different reward values are used to distinguish the actions according to their level of desirability, for example to signal to the agent that higher priority for failed tests is most desirable, followed by lower execution times for test cases with identical verdicts. We assign intermediary reward values, between 0 and 1, when ranking test cases with the same verdict, since the detection of failures is more important than saving computation resources. Nevertheless, due to the very small failure rate of the subjects (Section~\ref{sec:validation}) used in our experiments, we use a relatively large reward value (0.5) to signal the agent about the importance of ranking test case with the same verdicts, a frequent task when failures are rare. 

Overall, the action space and observation space of pairwise ranking are relatively small, which is expected to improve the scalability of RL. On the other hand, it is a priori unclear whether learning pairwise comparisons is sufficient to reach a good ranking accuracy. Indeed, to obtain a complete ranking we rely on sorting algorithms based on imperfect pairwise comparisons. 
We provide empirical results regarding the performance of this ranking model in Section~\ref{sec:validation}.

\subsection{Train an RL agent.}
\label{sec:traning}
We use state-of-the-art RL algorithms, listed in Table~\ref{tab:algos}, to train an agent. However, their applicability is limited by the type of their action space. More specifically, \textit{A2C}, \textit{PPO1}, \textit{PPO2}, and \textit{TRPO} are applied to all of the ranking models, \textit{DQN}, \textit{ACKTR}, and \textit{ACER} are applied only to pairwise and listwise ranking, and \textit{DDPG}, \textit{TD3}, and \textit{SAC} are applied only to pointwise ranking.

In the context of many systems, especially safety-critical systems, an RL agent cannot be trained online directly via interacting with a real environment. Instead, the agent is trained using a simulator that models the environment or replays the logs of system executions~\cite{dulac2019challenges}. In the latter case, after training an initial agent, the agent is deployed into the real environment. However, it is evaluated and trained offline based on new incoming logs in an iterative manner. Such offline training fits the CI context in which the execution logs of test cases are available and can be replayed to train an RL agent at the end of each CI cycle. 

Regarding offline training, at the beginning, we train the agent based on available execution history and then, for each cycle, (1) we use the agent to rank test cases in the prediction mode (no training  policy update is done while predicting), (2) we apply the ranking and capture the execution logs, and (3) we use the new logs to train the agent in an offline mode that allows the agent to adapt to new changes. The last step is fast since only the logs of one cycle are used for training, thus enabling offline training after each cycle.

\subsection{Integration of an RL agent into CI Environments}
\label{sec:integrationRL}
\begin{figure}
    \centering
    \includegraphics[width=8cm]{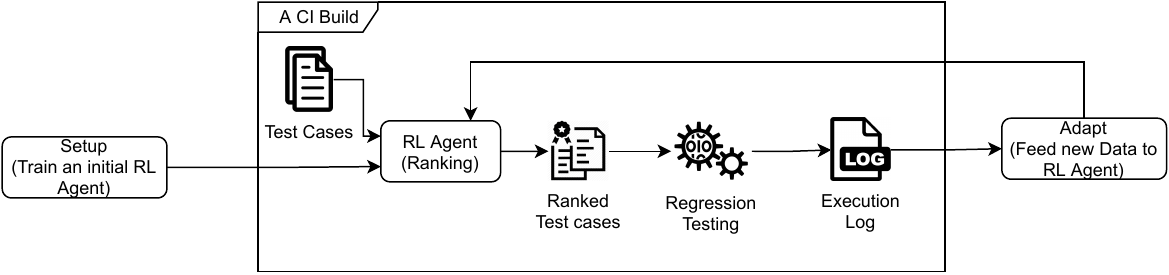}
    \caption{\color{black}Integrating an RL Agent in a CI Environment for Test Case Prioritization}
    \label{fig:IntegrateRL}
\end{figure}

\TR{E3, R2.2., R2.6, R3.6}{Figure~\ref{fig:IntegrateRL} shows how an RL agent can be integrated with the CI environment to prioritize regression test cases during a CI build. First, the agent needs to be trained based on the available data (test case execution history and related features extracted from the source code history) to reach a satisfactory accuracy. The trained agent is then deployed in the production environment and can be invoked by passing test case features in each CI cycle to rank test cases. Test cases are executed according to their ranks during regression testing, and their execution logs are captured. At pre-determined times or when the RL agent's accuracy is below a certain threshold, the execution logs are fed to the agent (i.e., logs are replayed) to adapt to the new changes. Both steps \textit{Setup} and \textit{Adapt} are done offline (i.e., are not done during a build process) via replaying the test cases execution logs, and therefore none of them delays the CI build. However, the ranking by the RL agent and calculation of test case features can delay the CI build. As we will discuss in Section~\ref{sec:validation}, the ranking time is negligible and the calculation time of test case features is in the order of seconds.
}
\section{Validation}
\label{sec:validation}
This section reports on the experiments we conducted to assess the accuracy and cost of the proposed RL configurations, and compare them with baselines. We first discuss the datasets of the study, evaluation metrics, comparison baselines, research questions, and our experimental setup. Then, we present the results and discuss their practical implications. The source code of our implementation and the results of experiments can be found here\footnote{\url{https://github.com/moji1/tp_rl}}.


\subsection{Datasets}
\label{sec:datasets}
\begin{table*}[t!]
\centering
	\caption{{\color{black} Data sets}} 
	\vspace{-.3cm}
	\label{tab:subjects}
    \begin{tabular}{cccccccc}
    \hline
     \small \textbf{Data set} &      \small \textbf{Type.}  &      \small \textbf{Cycles}  &      \small \textbf{Logs} &      \small \textbf{Fail Rate (\%)} & \small \textbf{Failed Cycles} &  \small \textbf{Avg. Calc. Time (Avg) Enriched Features ~\cite{bertolinolearning}}  \\ 
    \hline 

     \small \textbf{Paint-Control}   &\small   Simple&  332&  25,568 &  19.36\ & 252 & NA%
  \\

   \small \textbf{IOFROL}     & \small  Simple&  209 &  32,118 &  28.66\% & 203 & NA\\

 
        \small \textbf{Codec}     & \small   Enriched&   178 &  2,207 & 0\% & 0 & 1.78\\

        \small \textbf{Compress}     & \small   Enriched&   438 &  10,335 & 0.06\% & 7  & 3.64\\
   
        \small \textbf{Imaging}     & \small   Enriched&   147 &  4,482 & 0.04\% & 2& 5.60\\
   
        \small \textbf{IO}     & \small    Enriched&  176 &  4,985 & 0.06\% & 3 & 2.88\\
   
        \small \textbf{Lang}     & \small   Enriched&  301 &  10,884 & 0.01\%  & 2 & 5.58\\
  
        \small \textbf{Math}     & \small   Enriched &  55 &  3,822 & 0.01\% & 7. & 9.46\\

     \hline
    \end{tabular}

\end{table*}

We ran experiments on two categories of datasets: simple and enriched history datasets. The former consists of the execution history of two projects that were made publicly available by previous work \cite{spieker2017reinforcement}. As discussed in Definition ~\ref{def:testcasefeatures}, simple history data only contains the age, average execution time, and verdicts of test cases. Such datasets are representative of regression testing situations where source code is not available. Enriched datasets (six projects) consist of the augmented history data (execution history and code features) from the Apache Commons projects, which were made publicly accessible ~\cite{bertolinolearning}. The projects are written in Java and their build is managed using Maven. Enriched datasets represent testing situations where source code is available but full coverage analysis is not possible, due to the time constraints imposed by CI. 

Table~\ref{tab:subjects} lists the characteristics of the datasets. They contain the execution logs of $55$-to-$438$ CI cycles, each of which contains at least six test cases. We do not consider cycles with less test cases as (1) there is no benefit to applying prioritization on a few test cases, and, (2) more importantly, as we will discuss later (Section ~\ref{sec:metrics}), ranking a few test cases is not a challenging task and tends to inflate the accuracy results, as even random ranking can be suitable. The number of test case execution logs ranges from $2,207$ to $32,118$. \TR{R2.6}{Further, the failure rates and number of failed cycles (i.e., cycles that failed due to the failure of at least one regression test case, as defined in Def.~\ref{def:ci})} in enriched datasets are very low, ranging from 0 to 0.06 and 0 to 7, respectively, while the failure rates and number of failed cycles in simple datasets are abnormally high, ranging from $19.36$\% to $28.43$\% and $203$ to $252$, respectively.

\TR{R2.2, R2.6, R3.6}{Finally, the last column of Table~\ref{tab:subjects}, shows the average calculation time of enriched features per cycle (Def.~\ref{def:testcasefeatures}), based on the paper~\cite{bertolinolearning} that shared the enriched datasets. The calculation time ranges between 1.78 and 9.46 seconds per cycle across all datasets, which we consider a reasonable overhead in practice. }  

\subsection{Evaluation Metrics}
\label{sec:metrics}
We use two evaluation metrics to measure the accuracy of prioritization techniques, that are both used in the literature and are described below in turn.

\subsubsection{Normalized Rank Percentile Average (NRPA)}
We adopt the Normalized Rank Percentile Average (NRPA) ~\cite{bertolinolearning} for two reasons: (1) its capacity to measure the overall performance of a ranking, regardless of the context of the problem or the ranking criteria (e.g., fault detection for test prioritization), and (2) to be able to compare with the related work that uses \textit{NRPA} as the evaluation metric. \textit{NRPA} measures how close a predicted ranking of items is to the optimal ranking, i.e., the proportion of the optimal ranking that is contained in the predicted one. $NRPA$ ranges from 0 to 1, where higher values are preferable. Assuming a ranking algorithm $R$ that takes a set of $k$ items and generates an ordered sequence $s_e$, then $NRPA=\frac{RPA (s_e)}{RPA(s_0)}$, where $s_o$ is the optimal ranking of the items. Given any order of items (sequence $s$), 
$$
RPA (s) = \frac {\sum_{m \in s} \sum_{i=idx(s,m)}^{k} {|s|- idx(s_o,m)+1} }{k^2(k+1)/2}
$$
\indent where $idx(s,m)$ returns the position of $m$ in sequence $s$, and the lowest rank is given to an item with the highest priority.

\begin{figure}
    \centering
    \includegraphics[width=8cm]{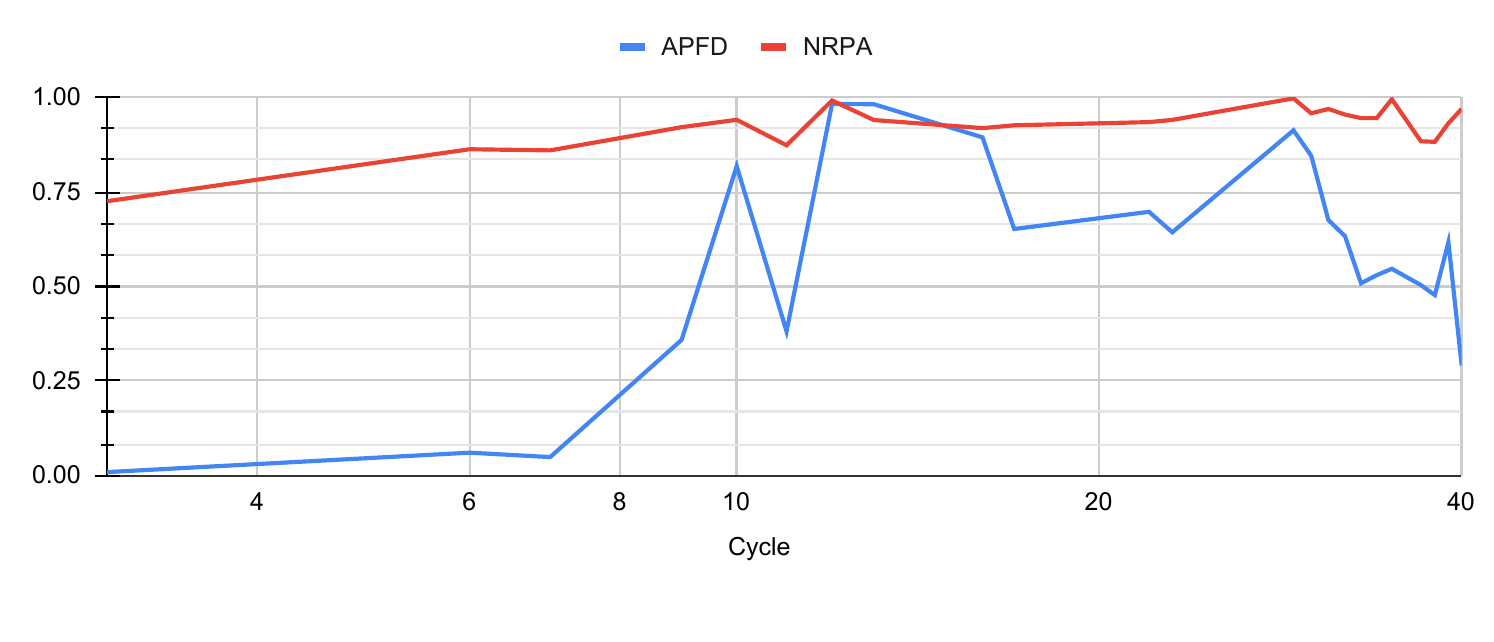}
    \caption{APFD versus NRPA for Algorithm A2C across the first 40 cycles of dataset Paint-Control }
    \label{fig:nrpavsapfd}
\end{figure}

\begin{figure}
    \centering
    \includegraphics[width=8cm]{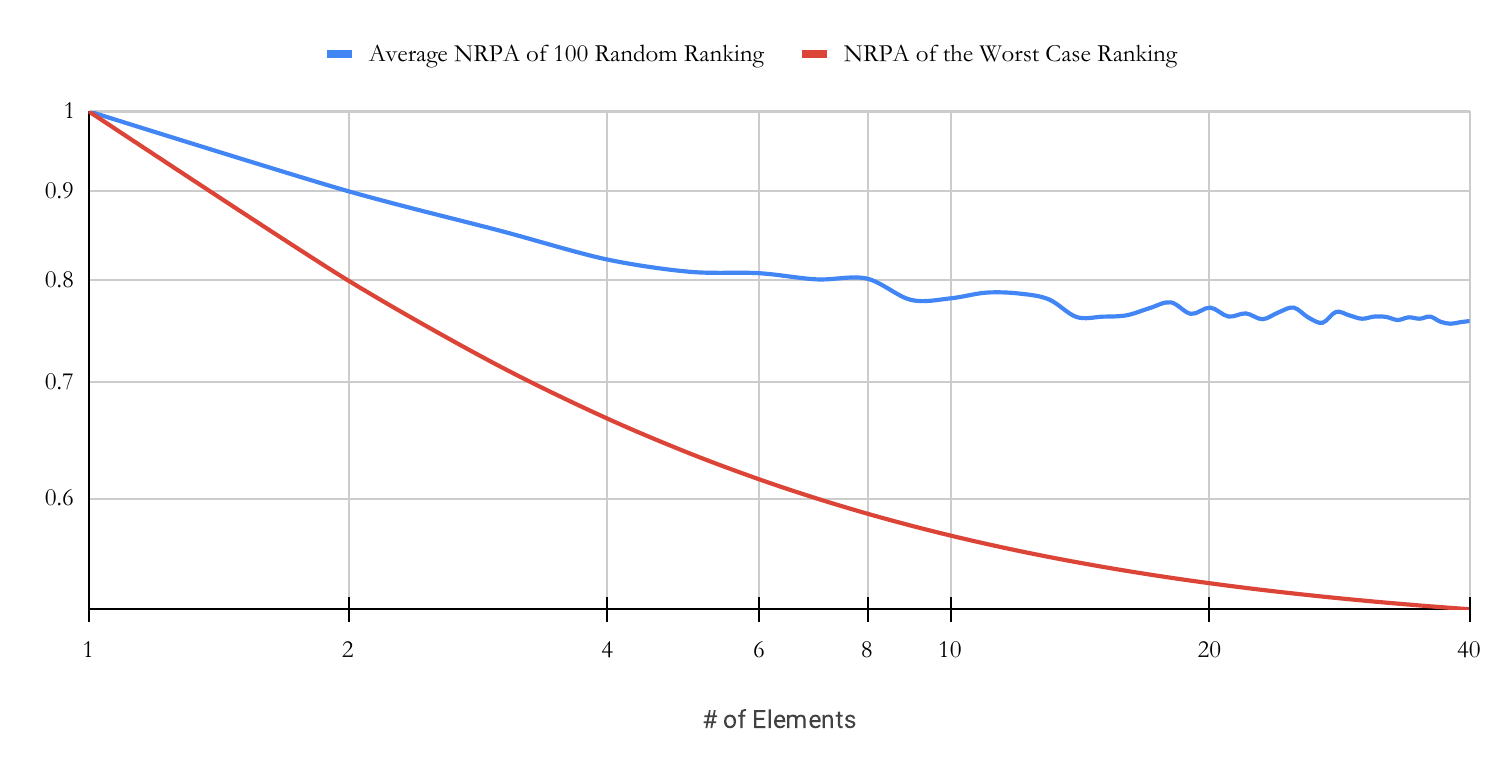}
    \caption{Worst and average NRPA values for 100 random rankings, across a range from 1 to 40 test cases (logarithmic scale)}
    \label{fig:nrpatrend}
\end{figure}

\subsubsection{Average Percentage of Faults Detected (APFD)}

While \textit{NRPA} is a suitable metric for measuring the accuracy of a ranking independently of the context, for regression testing, ranking failed test cases correctly (assigning them the highest priority) is much more important than ranking the rest of the test cases. However, \textit{NRPA} treats each test case equally regardless of their verdict and can be a misleading metric for test case prioritization in the presence of failures. For example, as shown in Figure~\ref{fig:nrpavsapfd} depicting the performance of the A2C algorithm using the pairwise model, for the \textit{Paint-Control} dataset in terms of APFD and NRPA. Algorithm A2C performs poorly in cycles 1-10, 13, and 40 in terms of prioritizing failed test cases, which is captured correctly by lower APFD values for these cycles. However, NRPA values contradict APFD values in all of these cycles (especially in cycle 40) because passing test cases are ranked properly based on their execution time and, therefore, they have high NRPA values. We observed such contradictory patterns in all of the datasets, thus suggesting that NRPA is not a good metric in the presence of failures, especially when only a small percentage of test cases fail per cycle. Therefore, we also make use of the well-known \textit{APFD} metric, since it measures how well a certain ranking can reveal faults early. 


\textit{APFD} measures the weighted average of the percentage of faults detected by the execution of test cases in a certain order. It ranges from 0 to 1, with higher numbers implying faster fault detection. The \textit{APFD} of an order $s_e$ is calculated as:
$$
APFD(s_e)= 1 - \frac {\sum_{t \in s_e}{idx(s_e,t)*t.v}} {|s_e|*m} + \frac{1}{2*|s_e|}
$$ 
where $m$ refers to the total number of faults.

While reviewing related work, we observed anomalies in the way APFD and NRPA were used and interpreted:  (1) they assumed an APFD value of 1 even when there were no failed test cases in a cycle, which led to misleading results, especially when the work reported the average APFD across all cycles, and (2) NRPA is reported for cycles with only a few test cases, even for cycles with one test case that always results in NRPA=1. To remedy these issues, (1) when there is no failed test case in a cycle, we do not report APFD and use NRPA as an alternative, and (2) we ignore all cycles with less than six test cases and do not report NRPA or APFD for them. To further justify our choice, in Figure~\ref{fig:nrpatrend}, we report the worst and average NRPA values obtained with random ranking for a range of items. For five or less, these values are high ($>$.60 and $>$.80, respectively) and, therefore, including such NRPA values in the evaluation would unrealistically boost the results.



Also, it is worth mentioning that while reaching the optimal ranking is challenging (NRPA=1), finding the worst ranking, in which no test case is ranked correctly (NRPA=0), is also difficult. Therefore, the value of NRPA can be relatively high, even for random ranking, and needs to be interpreted carefully.

\subsection{Comparison Baselines}
\label{sec:baselines}
Two papers report the use of RL for test case prioritization in the context of CI. The first~\cite{spieker2017reinforcement} (RL-BS1) applies RL on three simple history datasets. Since the source code and data was made publicly available, we use this work as the first RL baseline (RL-BS1) and compare our suggested RL strategies with the best RL configuration from that work, based on the two out of three simple datasets used by the paper (the first two datasets in Table~\ref{tab:subjects}). We left one dataset out, the Google Shared dataset of Test suite Results (GSDTR) that was originally provided by Elbaum et. al~\cite{elbaum2014techniques}. GSDTR contains a sample of 3.5 Million test suite execution results from a fast and large scale continuous testing infrastructure of several google products, developed using different programming languages. Unfortunately, RL-BS1 treats the entire dataset as if it were the CI execution logs of one product, which is not correct. Since the main focus, in most of ML-based test case prioritization work, including this work and RL-BS1, is creation of a ML model for a specific product based on its CI logs, using this dataset is not an option. Note that we have tried to divide the dataset into separate product-specific datasets but product information is missing. 


Second, the most recent related work by Bertolino et al. ~\cite{bertolinolearning} applies three different implementations (i.e., Shallow Network, Deep Neural Network, Random Forest) of RL on enriched datasets (the last six datasets). The datasets, implementations, and detailed results of these experiments were also made publicly available and we also use the best configuration of this work as a second baseline (RL-BS2) to compare our work based on the six enriched datasets.

Note, however, that our reported results deviates from published results due to the issues we considered to be inappropriate usage of APFD and NRPA, as discussed in Section ~\ref{sec:metrics}. Further, the study included cycles with less than five test cases, as discussed in Section ~\ref{sec:metrics}, thus matching situations where prioritization is not needed and also resulting in inflated average APFD and NRPA values. Last, we applied each of the baselines only on the datasets for which they were originally used, because their implementation does not support the use of other datasets, thus requiring changes which might introduce errors.



While the primary motivation of the use of RL is in dealing with the dynamic nature of test prioritization in a CI context, we still would like to compare RL performance with the best supervised learning technique. We would like to determine whether we can benefit from the practical advantages of RL (adaptation) without losing significant accuracy in ranking compared with supervised learning, as reported by Bertolino et al.~\cite{bertolinolearning}. Thus, we use the best ranking supervised learning technique (MART), based on reported results, as the third baseline (MART) of comparison. As discussed in Section~\ref{sec:introdution}, MART does not support incremental learning~\cite{zhang2019incremental}, a feature that is essential to cope with frequently-changing CI environments.



{
\subsection{Research Questions}
\begin{itemize}
   
 \item \textbf{RQ1} 
 How do the selected state-of-the-art RL solutions (Table~\ref{tab:algos}) perform in terms of prioritization accuracy and cost, using the simple and enriched datasets?
 \begin{itemize}
  \item \textbf{RQ1.1} Which combinations of RL algorithms and ranking models perform better? 
  \item \textbf{RQ1.2} Which of the three ranking models (pointwise, pairwise, and listwise) perform better across RL algorithms? 
    \item \textbf{RQ1.3} Which of the RL algorithms perform better across ranking models?

 \end{itemize}

  \item \textbf{RQ2} 
    How does the best RL solution identified in RQ1 perform w.r.t the comparison baselines, based on their respective original datasets (Section ~\ref{sec:baselines})? 
 \begin{itemize}
  \item \textbf{RQ1.1} How does the best RL solution perform compared to the RL baselines (RL-BS)? 
    \item \textbf{RQ2.2} How does the best RL solution perform compared to the ML baseline (MART)?
 \end{itemize}

\end{itemize}

\subsection{Experimental Setup and Configurations}
\TR{R3.5}{There are three ranking models from information retrieval. Thus to make our study comprehensive, we implemented\footnote{\url{https://github.com/moji1/tp_rl}} the three ranking models using state-of-the-art RL algorithms, as discussed in Section ~\ref{sec:traning}. We have used the Gym library~\cite{Gymlibrary} to simulate the CI environment using execution logs and relied on the implementation of RL algorithms provided by the Stable Baselines (v2.10.0) \cite{stable-baselines}. For each dataset, we ran three experiments corresponding to the three ranking models:  pairwise, listwise, and pointwise. As discussed in Section~\ref{sec:traning}, the applicability of RL algorithms is limited by the type of their action space. Thus, pairwise and listwise models involve seven experiments for each dataset, one per each RL technique that can support discrete action spaces (i.e., A2C, TRPO, PPO1-2, ACER, ACKTR, and DQN). Similarly,  pointwise involve eight experiments for each dataset using RL techniques that support continuous action spaces (i.e., A2C, TRPO, PPO1-2, SAC, DDPG, ACKTR, and TD3).} During the experiments, we observed that training using ACKTR with listwise ranking is extremely slow (more than 12 hours for a cycle), which makes it inappropriate for this work. Thus we dropped experiments using ACKTR and listwise ranking. The total number of experiments is therefore 168, during each of which an agent is trained for each cycle, and then tested on the subsequent cycles. Overall, this resulted into 308,448 (number of all cycles * 168 experiments) RL agent training and evaluation instances.  
The process of agent training and evaluation is incremental as it is started by training an agent by replaying the execution logs of the first cycle, followed by the evaluation of the trained agent on the second cycle, which is then followed by replaying logs of the second cycle to improve the agent, and so on.  

To ensure that enough training takes place, we used the minimum of $200* n*log_2(n)$ (corresponding to 200 episodes for each pairwise training instance) and one million steps for each training instance (training for each cycle), where $n$ refers to the number of test cases of the cycle. We stop when we either reach the set budget of steps per training instance or when the episode reward (sum of rewards across the steps of an episode) cannot be improved for more than 100 consecutive episodes (i.e., when the agent reaches plateau). The number of steps above is set based on the worst-case scenario, i.e., the pairwise training, in which an episode of training requires $n*log_2(n)$ steps. Often, RL algorithms come with a set of parameters that can be tuned. However, we use default parameters for all the experiments and leave parameter optimization to future experiments. Each experiment was run once using the HPC facilities of the University of Luxembourg~\cite{VBCG_HPCS14} with the same configuration of 3 CPU cores, and 20 GiB memory. During the experiments, the rank of each test, along with the required time for training and evaluation, were recorded to answer the RQs. Note that, even though each of the experiments is run once due to the massive computation time (more than 46 days with three CPU cores, and 20 GiB memory) required by all the experiments, our analysis is based on many cycles (see Table~\ref{tab:subjects}) and 308K training and evaluation instances, which allows us to account for randomness in RL algorithms and draw safe conclusions.


\begin{table*}[t]
\caption{The average performance of different configurations in terms of  APFD and NRPA, along with the results of the three baselines (Section ~\ref{sec:baselines}). The index in each cell shows the position of a configuration (row) with respect to others for each dataset (column) in terms of NRPA or APFD, based on statistical testing.}
\label{tab:results}
\begin{center}
\begin{tabular}{ l|cccccccccc} 
\hline
\multicolumn{1}{c|}{\textbf{}} & \multicolumn{1}{c}{\small \textbf{RM}} & 
\multicolumn{1}{c}{\small \textbf{IOFROL}} & 
\multicolumn{1}{c}{\small \textbf{Paint.} }& 
\multicolumn{1}{c}{\small \textbf{CODEC}} & 
\multicolumn{1}{c}{\small \textbf{IMAG}} & 
\multicolumn{1}{c}{\small \textbf{IO}} & 
\multicolumn{1}{c}{\small \textbf{COMP}} & 
\multicolumn{1}{c}{\small \textbf{LANG}} &
\multicolumn{1}{c}{\small \textbf{MATH}}  
\\
& & 
\small (APFD) &  
\small (APFD) & 

\small (NRPA)  & 
\small (NRPA)  & 
\small (NRPA)  & 
\small (NRPA)  & 
\small (NRPA) &  
\small (NRPA)   
\\
\hline
\hline
\multirow{3}{.40cm}{\small \textbf{A2C}} & \small \textbf{PA} & \small  .55$\pm$.13 \circled{4} & \small  .72$\pm$.24 \circled{1} & \small  .97$\pm$.04 \circled{1} & \small  .96$\pm$.05 \circled{1} & \small  .98$\pm$.02 \circled{1} & \small  .98$\pm$.03 \circled{2} & \small  .95$\pm$.04 \circled{2} & \small  .96$\pm$.04 \circled{1} \\ 
& \small \textbf{PO} & \small  .52$\pm$.14 \circled{5} & \small  .57$\pm$.23 \circled{5} & \small  .89$\pm$.07 \circled{6} & \small  .92$\pm$.05 \circled{8} & \small  .91$\pm$.05 \circled{3} & \small  .92$\pm$.05 \circled{6} & \small  .86$\pm$.07 \circled{5} & \small  .90$\pm$.04 \circled{2} \\ 
& \small \textbf{LI} & \small  .50$\pm$.10 \circled{9} & \small  .48$\pm$.18 \circled{9} & \small  .78$\pm$.08 \circled{8} & \small  .77$\pm$.06 \circled{12} & \small  .76$\pm$.05 \circled{4} & \small  .76$\pm$.06 \circled{9} & \small  .77$\pm$.05 \circled{8} & \small  .76$\pm$.06 \circled{5} \\ 
\hline 
\multirow{2}{.40cm}{\small \textbf{ACER}} & \small \textbf{PA} & \small  .56$\pm$.14 \circled{3} & \small  .73$\pm$.22 \circled{1} & \small  .98$\pm$.03 \circled{1} & \small  .96$\pm$.06 \circled{1} & \small  .98$\pm$.02 \circled{1} & \small  .98$\pm$.02 \circled{1} & \small  .96$\pm$.03 \circled{1} & \small  .96$\pm$.04 \circled{1} \\ 
& \small \textbf{LI} & \small  .50$\pm$.10 \circled{9} & \small  .48$\pm$.19 \circled{9} & \small  .79$\pm$.07 \circled{8} & \small  .77$\pm$.06 \circled{12} & \small  .76$\pm$.05 \circled{4} & \small  .77$\pm$.05 \circled{9} & \small  .77$\pm$.05 \circled{8} & \small  .77$\pm$.05 \circled{5} \\ 
\hline 
\multirow{2}{.40cm}{\small \textbf{ACKTR}} & \small \textbf{PA} & \small  .57$\pm$.13 \circled{1} & \small  .68$\pm$.22 \circled{2} & \small  .93$\pm$.09 \circled{5} & \small  .94$\pm$.07 \circled{5} & \small  .77$\pm$.05 \circled{4} & \small  .97$\pm$.03 \circled{4} & \small  .95$\pm$.04 \circled{3} & \small  .95$\pm$.05 \circled{1} \\ 
& \small \textbf{PO} & \small  .52$\pm$.14 \circled{5} & \small  .57$\pm$.24 \circled{5} & \small  .77$\pm$.08 \circled{8} & \small  .78$\pm$.08 \circled{11} & \small  .74$\pm$.08 \circled{4} & \small  .77$\pm$.06 \circled{9} & \small  .78$\pm$.06 \circled{7} & \small  .78$\pm$.06 \circled{5} \\ 
\hline 
\multirow{1}{.40cm}{\small \textbf{DDPG}} & \small \textbf{PO} & \small  .52$\pm$.13 \circled{5} & \small  .62$\pm$.22 \circled{4} & \small  .88$\pm$.08 \circled{7} & \small  .82$\pm$.07 \circled{10} & \small  .87$\pm$.09 \circled{3} & \small  .82$\pm$.07 \circled{8} & \small  .80$\pm$.07 \circled{6} & \small  .86$\pm$.07 \circled{3} \\ 
\hline 
\multirow{2}{.40cm}{\small \textbf{DQN}} & \small \textbf{PA} & \small  .53$\pm$.13 \circled{5} & \small  .67$\pm$.23 \circled{3} & \small  .94$\pm$.06 \circled{3} & \small  .95$\pm$.06 \circled{5} & \small  .98$\pm$.02 \circled{1} & \small  .97$\pm$.03 \circled{3} & \small  .95$\pm$.04 \circled{2} & \small  .94$\pm$.05 \circled{1} \\ 
& \small \textbf{LI} & \small  .50$\pm$.10 \circled{8} & \small  .50$\pm$.19 \circled{8} & \small  .79$\pm$.07 \circled{8} & \small  .77$\pm$.06 \circled{12} & \small  .76$\pm$.05 \circled{4} & \small  .77$\pm$.05 \circled{9} & \small  .77$\pm$.05 \circled{8} & \small  .76$\pm$.05 \circled{5} \\ 
\hline 
\multirow{3}{.40cm}{\small \textbf{PPO1}} & \small \textbf{PA} & \small  .56$\pm$.14 \circled{3} & \small  .72$\pm$.24 \circled{1} & \small  .97$\pm$.04 \circled{2} & \small  .96$\pm$.05 \circled{3} & \small  .98$\pm$.02 \circled{1} & \small  .98$\pm$.03 \circled{2} & \small  .95$\pm$.04 \circled{2} & \small  .96$\pm$.04 \circled{1} \\ 
& \small \textbf{PO} & \small  .52$\pm$.14 \circled{5} & \small  .58$\pm$.24 \circled{5} & \small  .89$\pm$.09 \circled{6} & \small  .93$\pm$.05 \circled{7} & \small  .90$\pm$.05 \circled{3} & \small  .90$\pm$.05 \circled{7} & \small  .86$\pm$.06 \circled{5} & \small  .84$\pm$.07 \circled{4} \\ 
& \small \textbf{LI} & \small  .51$\pm$.11 \circled{7} & \small  .56$\pm$.23 \circled{7} & \small  .79$\pm$.07 \circled{8} & \small  .77$\pm$.05 \circled{12} & \small  .76$\pm$.06 \circled{4} & \small  .77$\pm$.06 \circled{9} & \small  .78$\pm$.06 \circled{7} & \small  .78$\pm$.06 \circled{5} \\ 
\hline 
\multirow{3}{.40cm}{\small \textbf{PPO2}} & \small \textbf{PA} & \small  .57$\pm$.13 \circled{2} & \small  .71$\pm$.23 \circled{2} & \small  .97$\pm$.04 \circled{1} & \small  .96$\pm$.05 \circled{2} & \small  .98$\pm$.02 \circled{1} & \small  .98$\pm$.02 \circled{2} & \small  .96$\pm$.03 \circled{2} & \small  .96$\pm$.04 \circled{1} \\ 
& \small \textbf{PO} & \small  .52$\pm$.14 \circled{5} & \small  .57$\pm$.24 \circled{5} & \small  .93$\pm$.06 \circled{4} & \small  .93$\pm$.05 \circled{6} & \small  .95$\pm$.04 \circled{2} & \small  .94$\pm$.04 \circled{5} & \small  .89$\pm$.05 \circled{4} & \small  .85$\pm$.06 \circled{4} \\ 
& \small \textbf{LI} & \small  .51$\pm$.10 \circled{6} & \small  .49$\pm$.22 \circled{8} & \small  .79$\pm$.08 \circled{8} & \small  .78$\pm$.06 \circled{11} & \small  .76$\pm$.06 \circled{4} & \small  .77$\pm$.05 \circled{9} & \small  .78$\pm$.06 \circled{8} & \small  .77$\pm$.07 \circled{5} \\ 
\hline 
\multirow{1}{.40cm}{\small \textbf{SAC}} & \small \textbf{PO} & \small  .52$\pm$.14 \circled{5} & \small  .57$\pm$.24 \circled{6} & \small  .78$\pm$.09 \circled{8} & \small  .76$\pm$.07 \circled{12} & \small  .75$\pm$.08 \circled{4} & \small  .76$\pm$.08 \circled{9} & \small  .77$\pm$.07 \circled{8} & \small  .79$\pm$.07 \circled{5} \\ 
\hline 
\multirow{1}{.40cm}{\small \textbf{TD3}} & \small \textbf{PO} & \small  .52$\pm$.14 \circled{5} & \small  .58$\pm$.24 \circled{5} & \small  .78$\pm$.09 \circled{8} & \small  .78$\pm$.07 \circled{11} & \small  .75$\pm$.08 \circled{4} & \small  .77$\pm$.07 \circled{9} & \small  .77$\pm$.07 \circled{8} & \small  .76$\pm$.07 \circled{5} \\ 
\hline 
\multirow{3}{.40cm}{\small \textbf{TRPO}} & \small \textbf{PA} & \small  .57$\pm$.13 \circled{2} & \small  .71$\pm$.24 \circled{1} & \small  .96$\pm$.04 \circled{2} & \small  .95$\pm$.07 \circled{4} & \small  .98$\pm$.03 \circled{1} & \small  .98$\pm$.02 \circled{2} & \small  .95$\pm$.04 \circled{2} & \small  .95$\pm$.05 \circled{1} \\ 
& \small \textbf{PO} & \small  .52$\pm$.14 \circled{5} & \small  .57$\pm$.23 \circled{5} & \small  .90$\pm$.07 \circled{6} & \small  .92$\pm$.05 \circled{9} & \small  .94$\pm$.04 \circled{2} & \small  .92$\pm$.04 \circled{6} & \small  .90$\pm$.06 \circled{4} & \small  .86$\pm$.06 \circled{3} \\ 
& \small \textbf{LI} & \small  .50$\pm$.11 \circled{8} & \small  .48$\pm$.19 \circled{9} & \small  .80$\pm$.08 \circled{8} & \small  .76$\pm$.05 \circled{12} & \small  .77$\pm$.07 \circled{4} & \small  .76$\pm$.06 \circled{9} & \small  .77$\pm$.06 \circled{8} & \small  .78$\pm$.06 \circled{5} \\ 
\hline 

\hline

\small \textbf{Optimal} & \small \textbf{NA} & \small  .79$\pm$.14 & \small  .89$\pm$.14  & NA & NA & NA & NA & NA  & NA \\ 
\hline

\small \textbf{RL-BS1} & \small \textbf{PO} & \small  .63$\pm$.16 & \small  .74$\pm$.24  & NA & NA & NA & NA & NA  & NA \\ 
\hline
\small \textbf{RL-BS2} & \small \textbf{PO} & NA & NA & \small .90$\pm$.05 &  \small .89$\pm$.09 &  \small .84$\pm$.13 &  \small .90$\pm$.05 & \small .89$\pm$.07 &  \small .84$\pm$.13 \\ 
\hline

\small \textbf{MART}   & \small \textbf{PR} &  NA & NA & \small .96$\pm$.03 & \small .90$\pm$.05 & \small  .93$\pm$.02 & \small .96$\pm$.02 &  \small .94$\pm$.04  &  \small .95$\pm$.02 \\ 
\hline
\hline

\multicolumn{9}{l}{\footnotesize \textit{PR}: Pairwise, \textit{PO}: Pointwise, \textit{LI}: Listwise}

\end{tabular}
\end{center}

\end{table*}

\begin{table}[t!]
\centering
	\caption{Common Language Effect Size between one of the Worst and Best Configurations for each Data Set based on Accuracy} 
	\vspace{-.3cm}
	\label{tab:cle}
    \begin{tabular}{ccccc}
    \hline
     \small \textbf{Data set} &      \small \textbf{Best Conf.}  &      \small \textbf{Worst Conf.}  &      \small \textbf{CLE} \\ 
    \hline

\small \textbf{IOFROL} &\small PAIRWISE-ACKTR & LISTWISE-ACER  & .701 \\ 
\small \textbf{Paint.} &\small PAIRWISE-A2C & LISTWISE-TRPO  & .786 \\ 
\small \textbf{CODEC} &\small PAIRWISE-PPO2 & LISTWISE-TRPO  & .973 \\ 
\small \textbf{IMAG} &\small PAIRWISE-A2C & LISTWISE-TRPO  & .981 \\ 
\small \textbf{IO} &\small PAIRWISE-DQN & LISTWISE-TRPO  & .999 \\ 
\small \textbf{COMP} &\small PAIRWISE-ACER & LISTWISE-TRPO  & .997 \\ 
\small \textbf{LANG} &\small PAIRWISE-ACER & LISTWISE-TRPO  & .986 \\ 
\small \textbf{MATH} &\small PAIRWISE-DQN & LISTWISE-TRPO  & .963 \\

     \hline
    \end{tabular}

\end{table}


\begin{table*}[t]
\caption{The sum of training time (Minutes) for all cycles across datasets and configurations.}
\label{tab:results-training}
\begin{center}
\begin{tabular}{ p{1.1cm}|cccccccccccccccccc} 
\hline
& {\small \textbf{RM}} &  
{\small \textbf{IOFROL}} & 
{\small \textbf{Paint.} }& 
{\small \textbf{CODEC}} & 
{\small \textbf{IMAG}} & 
{\small \textbf{IO}} & 
{\small \textbf{COMP}} & 
{\small \textbf{LANG}} &
{\small \textbf{MATH}}  

\\
\hline
\hline

\multirow{3}{.40cm}{\small \textbf{A2C}} & \small \textbf{PA} & \small  911.8 \circled{2} & \small  713.5 \circled{3} & \small  187.4 \circled{2} & \small  213.0 \circled{9} & \small  253.2 \circled{10} & \small  986.0 \circled{4} & \small  602.3 \circled{4} & \small  107.4 \circled{4} \\ 
& \small \textbf{PO} & \small  1047.7 \circled{3} & \small  737.5 \circled{3} & \small  227.0 \circled{6} & \small  204.9 \circled{8} & \small  237.3 \circled{9} & \small  1724.4 \circled{15} & \small  1026.0 \circled{11} & \small  149.3 \circled{4} \\ 
& \small \textbf{LI} & \small  1718.3 \circled{5} & \small  775.1 \circled{4} & \small  209.1 \circled{5} & \small  181.0 \circled{7} & \small  26.3 \circled{11} & \small  1142.8 \circled{10} & \small  545.4 \circled{3} & \small  222.9 \circled{4} \\ 
\hline 
\multirow{2}{.40cm}{\small \textbf{ACER}} & \small \textbf{PA} & \small  944.6 \circled{3} & \small  715.6 \circled{3} & \small  175.7 \circled{1} & \small  147.3 \circled{3} & \small  184.8 \circled{3} & \small  905.4 \circled{1} & \small  61.2 \circled{4} & \small  11.9 \circled{4} \\ 
& \small \textbf{LI} & \small  1983.7 \circled{6} & \small  74.7 \circled{3} & \small  223.4 \circled{6} & \small  155.7 \circled{4} & \small  196.4 \circled{4} & \small  101.2 \circled{5} & \small  71.7 \circled{7} & \small  254.0 \circled{4} \\ 
\hline 
\multirow{2}{.40cm}{\small \textbf{ACKTR}} & \small \textbf{PA} & \small  949.8 \circled{3} & \small  762.1 \circled{4} & \small  189.9 \circled{2} & \small  161.7 \circled{5} & \small  199.1 \circled{4} & \small  1012.9 \circled{5} & \small  635.1 \circled{5} & \small  127.3 \circled{4} \\ 
& \small \textbf{PO} & \small  1032.4 \circled{3} & \small  641.7 \circled{2} & \small  202.6 \circled{4} & \small  148.5 \circled{3} & \small  231.1 \circled{9} & \small  109.6 \circled{8} & \small  665.4 \circled{6} & \small  122.2 \circled{4} \\ 
\hline 
\multirow{1}{.40cm}{\small \textbf{DDPG}} & \small \textbf{PO} & \small  2576.0 \circled{7} & \small  1509.5 \circled{7} & \small  254.9 \circled{9} & \small  254.4 \circled{11} & \small  349.6 \circled{12} & \small  1133.3 \circled{10} & \small  985.5 \circled{10} & \small  221.6 \circled{4} \\ 
\hline 
\multirow{2}{.40cm}{\small \textbf{DQN}} & \small \textbf{PA} & \small  2021.0 \circled{6} & \small  1343.4 \circled{6} & \small  242.3 \circled{8} & \small  389.2 \circled{13} & \small  429.0 \circled{13} & \small  1277.8 \circled{13} & \small  91.7 \circled{10} & \small  236.5 \circled{4} \\ 
& \small \textbf{LI} & \small  10652.9 \circled{10} & \small  3602.9 \circled{10} & \small  638.1 \circled{12} & \small  564.8 \circled{14} & \small  629.3 \circled{15} & \small  2049.1 \circled{16} & \small  2024.2 \circled{14} & \small  1842.2 \circled{8} \\ 
\hline 
\multirow{3}{.40cm}{\small \textbf{PPO1}} & \small \textbf{PA} & \small  848.9 \circled{1} & \small  664.1 \circled{2} & \small  17.6 \circled{1} & \small  133.8 \circled{2} & \small  167.8 \circled{1} & \small  908.8 \circled{1} & \small  555.6 \circled{3} & \small  98.3 \circled{3} \\ 
& \small \textbf{PO} & \small  836.4 \circled{1} & \small  668.5 \circled{2} & \small  186.5 \circled{2} & \small  138.9 \circled{3} & \small  203.9 \circled{6} & \small  1102.9 \circled{9} & \small  845.4 \circled{9} & \small  72.4 \circled{1} \\ 
& \small \textbf{LI} & \small  5183.4 \circled{9} & \small  1194.5 \circled{6} & \small  226.9 \circled{6} & \small  225.3 \circled{10} & \small  273.5 \circled{11} & \small  1201.5 \circled{12} & \small  961.5 \circled{10} & \small  713.4 \circled{7} \\ 
\hline 
\multirow{3}{.40cm}{\small \textbf{PPO2}} & \small \textbf{PA} & \small  922.8 \circled{2} & \small  72.1 \circled{3} & \small  316.5 \circled{11} & \small  174.7 \circled{6} & \small  224.0 \circled{8} & \small  961.3 \circled{3} & \small  607.5 \circled{4} & \small  107.5 \circled{4} \\ 
& \small \textbf{PO} & \small  994.3 \circled{3} & \small  623.8 \circled{2} & \small  219.8 \circled{6} & \small  149.7 \circled{3} & \small  231.9 \circled{9} & \small  878.2 \circled{1} & \small  675.0 \circled{6} & \small  86.2 \circled{2} \\ 
& \small \textbf{LI} & \small  1726.0 \circled{4} & \small  731.5 \circled{3} & \small  195.0 \circled{3} & \small  165.1 \circled{5} & \small  243.8 \circled{9} & \small  1027.3 \circled{6} & \small  524.1 \circled{2} & \small  197.7 \circled{4} \\ 
\hline 
\multirow{1}{.40cm}{\small \textbf{SAC}} & \small \textbf{PO} & \small  3413.4 \circled{7} & \small  259.3 \circled{9} & \small  336.8 \circled{11} & \small  474.2 \circled{13} & \small  483.7 \circled{14} & \small  1592.4 \circled{15} & \small  1269.6 \circled{13} & \small  455.7 \circled{6} \\ 
\hline 
\multirow{1}{.40cm}{\small \textbf{TD3}} & \small \textbf{PO} & \small  3592.1 \circled{8} & \small  1821.2 \circled{8} & \small  291.3 \circled{10} & \small  37.1 \circled{12} & \small  414.3 \circled{13} & \small  1318.9 \circled{14} & \small  1086.5 \circled{12} & \small  333.8 \circled{5} \\ 
\hline 
\multirow{3}{.40cm}{\small \textbf{TRPO}} & \small \textbf{PA} & \small  758.3 \circled{1} & \small  611.1 \circled{1} & \small  189.5 \circled{2} & \small  13.5 \circled{1} & \small  176.4 \circled{2} & \small  927.6 \circled{2} & \small  556.8 \circled{3} & \small  89.4 \circled{2} \\ 
& \small \textbf{PO} & \small  806.9 \circled{1} & \small  626.3 \circled{2} & \small  207.2 \circled{5} & \small  135.0 \circled{2} & \small  201.6 \circled{5} & \small  1065.8 \circled{7} & \small  481.5 \circled{1} & \small  66.6 \circled{1} \\ 
& \small \textbf{LI} & \small  3005.5 \circled{7} & \small  855.5 \circled{5} & \small  217.6 \circled{7} & \small  161.1 \circled{5} & \small  215.1 \circled{7} & \small  1151.1 \circled{11} & \small  82.6 \circled{8} & \small  371.5 \circled{5} \\ 
\hline

\hline
\multicolumn{9}{l}{\footnotesize \textit{PR}: Pairwise, \textit{PO}: Pointwise, \textit{LI}: Listwise}

\end{tabular}
\end{center}

\end{table*}

\begin{table*}[t]
\caption{\color{black}The average of prediction (ranking) time (Seconds) for all cycles across datasets and configurations.}
\label{tab:results-prediction}
\begin{center}
\begin{tabular}{p{1.1cm}|cccccccccccccccccc} 
\hline
& {\small \textbf{RM}} &  
{\small \textbf{IOFROL}} & 
{\small \textbf{Paint.} }& 
{\small \textbf{CODEC}} & 
{\small \textbf{IMAG}} & 
{\small \textbf{IO}} & 
{\small \textbf{COMP}} & 
{\small \textbf{LANG}} &
{\small \textbf{MATH}}  

\\
\hline
\hline

\multirow{3}{.40cm}{\small \textbf{A2C}} & \small \textbf{PA} & \small  1.3$\pm$0.9\circled{7} & \small  1.0$\pm$0.2\circled{4} & \small  0.8$\pm$0.1\circled{1} & \small  0.8$\pm$0.1\circled{5} & \small  0.7$\pm$0.1\circled{1} & \small  0.8$\pm$0.1\circled{4} & \small  0.9$\pm$0.1\circled{3} & \small  1.1$\pm$0.5\circled{6} \\ 
& \small \textbf{PO} & \small  1.0$\pm$0.1\circled{3} & \small  0.8$\pm$0.1\circled{2} & \small  0.9$\pm$0.1\circled{4} & \small  0.6$\pm$0.1\circled{1} & \small  0.7$\pm$0.1\circled{1} & \small  0.6$\pm$0.0\circled{1} & \small  0.6$\pm$0.0\circled{1} & \small  0.7$\pm$0.1\circled{2} \\ 
& \small \textbf{LI} & \small  32$\pm$142\circled{6} & \small  4$\pm$14\circled{7} & \small  0.9$\pm$0.4\circled{3} & \small  1.4$\pm$2.9\circled{8} & \small  2.0$\pm$3.5\circled{9} & \small  4$\pm$25\circled{8} & \small  1.2$\pm$0.6\circled{8} & \small  5$\pm$20\circled{7} \\ 
\hline 
\multirow{2}{.40cm}{\small \textbf{ACER}} & \small \textbf{PA} & \small  2$\pm$0.9\circled{14} & \small  1.9$\pm$0.2\circled{10} & \small  1.8$\pm$0.1\circled{14} & \small  1.4$\pm$0.1\circled{13} & \small  1.4$\pm$0.1\circled{12} & \small  1.6$\pm$0.1\circled{14} & \small  1.8$\pm$0.1\circled{14} & \small  2.0$\pm$0.5\circled{13} \\ 
& \small \textbf{LI} & \small  11$\pm$26\circled{15} & \small  3$\pm$2\circled{12} & \small  2$\pm$0.2\circled{15} & \small  1.7$\pm$0.3\circled{15} & \small  1.7$\pm$0.4\circled{13} & \small  2.0$\pm$0.8\circled{16} & \small  3$\pm$1.3\circled{15} & \small  4$\pm$3\circled{14} \\ 
\hline 
\multirow{2}{.40cm}{\small \textbf{ACKTR}} & \small \textbf{PA} & \small  1.6$\pm$0.8\circled{11} & \small  1.4$\pm$0.2\circled{8} & \small  1.2$\pm$0.1\circled{7} & \small  1.0$\pm$0.1\circled{9} & \small  1.0$\pm$0.1\circled{4} & \small  1.2$\pm$0.1\circled{9} & \small  1.2$\pm$0.1\circled{10} & \small  1.4$\pm$0.5\circled{11} \\ 
& \small \textbf{PO} & \small  1.2$\pm$0.2\circled{5} & \small  1.0$\pm$0.1\circled{4} & \small  1.3$\pm$0.1\circled{9} & \small  1.0$\pm$0.1\circled{9} & \small  1.1$\pm$0.1\circled{7} & \small  1.2$\pm$0.1\circled{11} & \small  1.3$\pm$0.1\circled{11} & \small  1.2$\pm$0.1\circled{9} \\ 
\hline 
\multirow{1}{.40cm}{\small \textbf{DDPG}} & \small \textbf{PO} & \small  0.9$\pm$0.2\circled{2} & \small  0.9$\pm$0.1\circled{3} & \small  0.9$\pm$0.1\circled{2} & \small  0.7$\pm$0.1\circled{2} & \small  0.7$\pm$0.1\circled{1} & \small  0.7$\pm$0.1\circled{2} & \small  0.8$\pm$0.1\circled{2} & \small  0.7$\pm$0.2\circled{2} \\ 
\hline 
\multirow{2}{.40cm}{\small \textbf{DQN}} & \small \textbf{PA} & \small  1.7$\pm$1.0\circled{12} & \small  1.3$\pm$0.3\circled{8} & \small  1.2$\pm$0.3\circled{7} & \small  1.0$\pm$0.3\circled{10} & \small  1.0$\pm$0.3\circled{4} & \small  1.2$\pm$0.3\circled{10} & \small  1.2$\pm$0.2\circled{9} & \small  1.3$\pm$0.4\circled{10} \\ 
& \small \textbf{LI} & \small  9$\pm$44\circled{6} & \small  2.0$\pm$0.3\circled{11} & \small  1.5$\pm$0.6\circled{11} & \small  1.2$\pm$0.3\circled{11} & \small  1.1$\pm$0.2\circled{7} & \small  1.1$\pm$0.3\circled{10} & \small  1.3$\pm$0.4\circled{11} & \small  2$\pm$0.7\circled{15} \\ 
\hline 
\multirow{3}{.40cm}{\small \textbf{PPO1}} & \small \textbf{PA} & \small  1.4$\pm$1.0\circled{8} & \small  1.0$\pm$0.3\circled{5} & \small  0.8$\pm$0.2\circled{2} & \small  0.7$\pm$0.1\circled{3} & \small  0.7$\pm$0.2\circled{1} & \small  0.8$\pm$0.3\circled{5} & \small  0.9$\pm$0.2\circled{4} & \small  1.1$\pm$0.5\circled{6} \\ 
& \small \textbf{PO} & \small  0.8$\pm$0.1\circled{1} & \small  0.8$\pm$0.1\circled{2} & \small  0.8$\pm$0.2\circled{1} & \small  0.7$\pm$0.2\circled{3} & \small  0.7$\pm$0.1\circled{1} & \small  0.8$\pm$0.2\circled{4} & \small  0.7$\pm$0.2\circled{1} & \small  0.6$\pm$0.1\circled{1} \\ 
& \small \textbf{LI} & \small  406$\pm$405\circled{16} & \small  329$\pm$234\circled{15} & \small  3$\pm$6\circled{13} & \small  4$\pm$7\circled{16} & \small  19$\pm$72\circled{6} & \small  8$\pm$38\circled{8} & \small  62$\pm$173\circled{16} & \small  50$\pm$181\circled{7} \\ 
\hline 
\multirow{3}{.40cm}{\small \textbf{PPO2}} & \small \textbf{PA} & \small  1.5$\pm$1.0\circled{10} & \small  1.1$\pm$0.2\circled{6} & \small  0.9$\pm$0.1\circled{5} & \small  0.9$\pm$0.1\circled{7} & \small  1.0$\pm$0.1\circled{3} & \small  1.0$\pm$0.1\circled{7} & \small  1.0$\pm$0.1\circled{6} & \small  1.2$\pm$0.5\circled{8} \\ 
& \small \textbf{PO} & \small  1.0$\pm$0.1\circled{4} & \small  0.8$\pm$0.1\circled{1} & \small  1.0$\pm$0.1\circled{6} & \small  0.7$\pm$0.1\circled{4} & \small  0.9$\pm$0.1\circled{2} & \small  0.7$\pm$0.1\circled{3} & \small  0.9$\pm$0.1\circled{5} & \small  0.7$\pm$0.1\circled{3} \\ 
& \small \textbf{LI} & \small  396$\pm$462\circled{16} & \small  217$\pm$254\circled{14} & \small  4$\pm$24\circled{7} & \small  87$\pm$193\circled{17} & \small  14$\pm$47\circled{8} & \small  8$\pm$38\circled{8} & \small  32$\pm$104\circled{16} & \small  44$\pm$137\circled{7} \\ 
\hline 
\multirow{1}{.40cm}{\small \textbf{SAC}} & \small \textbf{PO} & \small  1.5$\pm$0.1\circled{9} & \small  1.2$\pm$0.1\circled{7} & \small  1.5$\pm$0.1\circled{12} & \small  1.2$\pm$0.1\circled{12} & \small  1.3$\pm$0.1\circled{10} & \small  1.2$\pm$0.1\circled{9} & \small  1.5$\pm$0.1\circled{13} & \small  1.1$\pm$0.1\circled{5} \\ 
\hline 
\multirow{1}{.40cm}{\small \textbf{TD3}} & \small \textbf{PO} & \small  1.0$\pm$0.1\circled{3} & \small  1.2$\pm$0.1\circled{7} & \small  1.2$\pm$0.1\circled{7} & \small  0.9$\pm$0.1\circled{6} & \small  1.0$\pm$0.1\circled{5} & \small  0.9$\pm$0.1\circled{6} & \small  1.1$\pm$0.1\circled{7} & \small  0.9$\pm$0.1\circled{4} \\ 
\hline 
\multirow{3}{.40cm}{\small \textbf{TRPO}} & \small \textbf{PA} & \small  1.9$\pm$0.9\circled{13} & \small  1.5$\pm$0.4\circled{9} & \small  1.4$\pm$0.3\circled{10} & \small  1.2$\pm$0.2\circled{11} & \small  1.2$\pm$0.4\circled{7} & \small  1.4$\pm$0.4\circled{12} & \small  1.4$\pm$0.3\circled{12} & \small  1.6$\pm$0.5\circled{12} \\ 
& \small \textbf{PO} & \small  1.3$\pm$0.3\circled{6} & \small  1.6$\pm$0.4\circled{9} & \small  1.6$\pm$0.5\circled{11} & \small  1.2$\pm$0.3\circled{12} & \small  1.4$\pm$0.3\circled{11} & \small  1.5$\pm$0.5\circled{13} & \small  1.2$\pm$0.2\circled{10} & \small  1.2$\pm$0.1\circled{9} \\ 
& \small \textbf{LI} & \small  245$\pm$409\circled{16} & \small  47$\pm$118\circled{13} & \small  1.2$\pm$0.3\circled{8} & \small  1.8$\pm$1.3\circled{14} & \small  2$\pm$4\circled{9} & \small  1.9$\pm$1.6\circled{15} & \small  158$\pm$287\circled{17} & \small  11$\pm$21\circled{7} \\ 
\hline 

\hline
\multicolumn{9}{l}{\footnotesize \textit{PR}: Pairwise, \textit{PO}: Pointwise, \textit{LI}: Listwise; Values greater than 2 are rounded}

\end{tabular}
\end{center}

\end{table*}

\subsection{Results and Discussion} 
\subsubsection{RQ1.}

\textbf{Overview}. Table~\ref{tab:results} shows the averages and standard deviations of APFD and NRPA for the eight datasets, using different configurations (combinations of ranking model and RL algorithm). Each column and row corresponds to a dataset and configuration, respectively. For example, the first column reports on how different configurations perform with \textit{Paint-Control}, and the first row shows how the combination of pointwise ranking and A2C works for all datasets. We use convention 
${[ranking\; model]-[RL \; algorithm]}$ to refer to configurations in the rest of the paper. For example \textit{pairwise-A2C} refers to a configuration of the pairwise ranking model and A2C algorithm.

\begin{definition} {(Relative Performance Rank)} \label{def:relativeperfrank}
For each dataset (column), the relative performance ranks of configurations in terms of APFD or NRPA are depicted with $\circled{n}$, where a lower rank $n$ indicates better performance. Assuming that $cnt$ of a configuration $cf$ denotes the number of configurations with a significantly lower average in terms of APFD or NRPA, a configuration with a higher $cnt$ is ranked lower, and two configurations with identical $cnt$ are ranked equal. For instance, for dataset \textit{IMAGE}, \TR{R1.3}{configurations \textit{pairwise-A2C} and \textit{pairwise-ACER} are ranked \circled{1},} because the $cnt$ of all three configurations are equal and significantly higher than other configurations.  
\end{definition}
 
To check for significant differences in results across configurations for a given dataset, we use Welch's ANOVA~\cite{welch1947generalization} to compare all configurations across all cycles, with one NRPA or APFD value per cycle. Then we perform the Games-Howell post-hoc test~\cite{games1976pairwise} to compare each pair of configurations. The significance level is set to 0.05, and therefore any difference with \textit{p}-value $<$= 0.05 is considered significant. 
We use Welch's ANOVA rather than one-way ANOVA because the variance in results across configurations are not equal. Also, we rely on the Games-Howell post-hoc test due to its compatibility with Welch's ANOVA (i.e., no assumption about equal variance). Using this post-hoc test also addresses the usual problems related to repeated testing when performing multiple comparisons (increased type-I error)~\cite{kim2014analysis}.

\TR{E3, R2.2, R2.6, R3.6}{
Table~\ref{tab:results-training} shows the overall training times for all experiments. Similarly, Table~\ref{tab:results-prediction} shows the averages and standard deviations of prediction (ranking) time for all experiments. Each column and row corresponds to a dataset and configuration, respectively. For each dataset (column), the relative performance ranking of configurations in terms of training/prediction time are depicted with $\circled{n}$, where a lower rank $n$ indicates smaller training/prediction time. The relative performance ranks are calculated according to Def.~\ref{def:relativeperfrank}, but based on training/prediction time. 
}

 \textbf{RQ1.1} As shown in Table \ref{tab:results}, multiple pairwise configurations perform best for some of the datasets but we cannot single out one configuration overall based on ranks. Configuration pairwise-ACER yields, however, the best averages. Also, based on the post-hoc test, pairwise-ACER perform best across all datasets except one, followed by pairwise-A2C and pairwise-PPO2. In contrast, listwise ranking with all algorithms, pointwise-TD3, and pointwise-SAC perform worst. \TR{R3.5}{As discussed in Sec.~\ref{sec:listwise}, we argue that due to the large observation space of listwise, it requires extensive training data, and under the same circumstances (i.e., same training data and steps of training) their accuracy can not be as good as the pairwise and pointwise configurations.}

 Also, for each dataset, we measured the effect size of the differences between configuration pairs based on Common Language Effect Size (CLE)~\cite{mcgraw1992common,arcuri2014hitchhiker}. CLE estimates the probability that a randomly sampled score from one population will be greater than a randomly sampled score from the other population. As shown in Table~\ref{tab:cle}, CLE values between one of the worst and best cases for the six enriched datasets are above 96\%, while they are 79\% and 70\%  for the simple datasets Paint-Control and IOFROL, respectively. Results therefore show the importance of selecting one of the best configurations. Also, relatively smaller CLEs for simple datasets suggests none of the configurations learned an adequate ranking strategy in these cases. This may not be surprising since learning an accurate policy for complex software systems cannot be expected to be always possible based on simple data.

\TR{E3, R2.2, R2.6, R3.6}{ 
 In terms of training time, as shown in Table~\ref{tab:results-training}, based once again on the post-hoc test, multiple configurations (pairwise or pointwise) perform well for some of the datasets. Still, we cannot recommend one specific configuration overall. Pairwise-TRPO and pairwise-PPO1 are the most efficient in terms of training time, followed by pairwise-ACER. In contrast, listwise and pairwise rankings with the DQN algorithm feature the worst training time. Also, pointwise-DDPG, pointwise-TD3, and pointwise-SAC are relatively slow across all datasets. As discussed in Section~\ref{sec:integrationRL}, since our approach uses offline training for the initial setup and adapting to new changes, training can therefore occur in the background without adding any delay to the CI build process. Thus, the differences in training time across configurations, which are in the order of minutes, do not constitute a practical issue. Also, the training in pointwise and pairwise modes could be made more efficient by relying on sampling to only replay a random and small subset of test execution logs during training. 
 }

\TR{E3, R2.2, R2.6, R3.6}{ 
 In terms of prediction time, as shown in Table~\ref{tab:results-prediction}, based once again on the post-hoc test, multiple pointwise configurations perform well for some of the datasets. Still, we cannot recommend one specific configuration overall. Pointwise-PPO1 and pointwise-A2C are the most efficient in terms of prediction time, which are followed by pointwise-A2C. In contrast, listwise ranking with the ACER, PPO1, PPO2, and TRPO algorithms feature the worst prediction times. The prediction times of pointwise and pairwise configurations do not exceed 2.22 seconds across all datasets and are therefore negligible overheads for CI builds. Thus, considering the low data collection time for enriched datasets ($<$ 10 seconds, last column of Table~\ref{tab:subjects}), it is safe to conclude that pairwise and pointwise approaches are practical choices in terms of computation overhead. In contrast, in several cases, prediction times of listwise configurations spike to more than 200 seconds and may not be applicable in practice.} \TR{R3.5}{A possible reason is related to how listwise ranking is modeled in Algorithm~\ref{code:listwise}, in which dummy test cases are used for padding to prevent test cases from being selected repeatedly. Thus, when the agent cannot learn a suitable policy, it selects dummy test cases many times, and that increases prediction time.
 }

\textbf{RQ1.2} As discussed above, pairwise ranking configurations fare relatively better than pointwise and listwise ranking in terms of accuracy (NRPA and APFD). Since applicable algorithms differ for each ranking model, such superior performance may therefore be due to the RL algorithms rather than the pairwise ranking model. Thus, to investigate this hypothesis, we perform four sets of Welch ANOVA and Games-Howell post-hoc tests based on the results of each of the four algorithms (A2C, PPO1, PPO2, TRPO) shared across ranking models and all enriched datasets. For each such algorithm, we create three sample groups corresponding to three ranking models based on the results of the algorithms for enriched datasets. We repeat the same analysis for the simple datasets too. The results show that, for enriched and simple datasets, there is a statistically significant difference for each algorithm across ranking models. Further, in all cases, regardless of the algorithm, pairwise fares better than pointwise and listwise. Similarly, pointwise fares better than listwise  in all cases, except for PPO1 where they are comparable.  

\TR{R3.5}{As discussed above (RQ1.1), due to the large observation space, the accuracy of listwise configurations is relatively lower than pointwise and pairwise. We also conjecture that in pairwise configurations, using a pair of test cases allows the agent to be trained on richer feature sets (see Section~\ref{sec:pairwise}) compared to the pointwise configurations that use a set of features based on a single test case (see Section~\ref{sec:pointwise}). In general, a ranking model that is trained based on pointwise features can be coarse due to the limited information captured by single document features~\cite{kang2011learning}. This is why state-of-the-art ranking models tend to use pairwise ranking~\cite{burges2010ranknet} (e.g., RankNet, LambdaRank, and LambdaMART).}


Concerning training time, again, we performed four sets of Welch ANOVA and Games-Howell post-hoc test for enriched and simple datasets, as discussed above but based on training time. The results show that, for enriched datasets, there is a statistically significant difference for each algorithm across ranking models. Further, in all cases, regardless of the algorithm, listwise fares worse than pointwise and pairwise, except A2C for which pointwise fares worst. Also, pairwise fares better than other models except for TRPO for which pairwise and pointwise are similar. Similarly, for simple datasets, listwise is worse than other models. Pairwise and pointwise have similar training time in all cases. 
As discussed in RQ1.1, however, the differences in training time across configurations, which are in the order of minutes, do not constitute a practical issue in our context.


\TR{E3, R2.2, R2.6, R3.6}{
Again, we performed four sets of Welch ANOVA and Games-Howell post-hoc tests for enriched and simple datasets, as discussed above but based on prediction time. The results show that there is a statistically significant difference for each algorithm across ranking models, for both enriched and simple datasets. Further, in all cases, regardless of the algorithm, listwise fares worse than pointwise and pairwise. Also, pointwise fares better than pairwise. 
As discussed in RQ1.1, the high prediction time of listwise ranking can entail practical issues in our context. However, the differences in prediction time between pairwise and pointwise across configurations are less than 2.22 seconds and do not have practical implications. 
}

\textbf{RQ1.3}
To analyze the relative accuracy of RL algorithms, we perform three sets of Welch ANOVA and Games-Howell post-hoc tests corresponding to the three ranking models, based on the result of all algorithms, across enriched datasets. Then we repeat the same analysis for simple datasets. The result shows that there are no significant differences between RL algorithms for enriched datasets when using the listwise ranking model. Similarly, there are no significant differences for simple datasets when using pointwise ranking. The result for the remaining cases are different as described next, assuming that where $>$ and =  denote greater and equal performance rank, respectively, as calculated in Definition ~\ref{def:relativeperfrank}.

\begin{itemize}

 \item \textit{Pairwise and enriched datasets:} ACER $>$ PPO2 $=$ A2C $=$  PPO1 $>$ TRPO $>$  DQN $>$ ACKTR 
 \item  \textit{Pointwise and enriched datasets:} PPO2 $>$ TRPO $>$ A2C $>$  PPO1 $>$ DDPG $>$  SAC =  TD3 =  ACKTR. 
 \item  \textit{Pairwise and simple datasets:} ACER $>$ PPO2 $=$ A2C $=$  PPO1 $>$ TRPO $=$ ACKTR $>$ DQN  
 \item  \textit{Listwise and simple datasets:} PPO1 $>$ PPO2 $=$ A2C $=$ PPO1 $=$ TRPO $=$ ACER $=$ DQN

\end{itemize}


To compare training time, we perform an analysis similar to the one above but based on training time at each cycle. Results clearly show that there is a significant difference across the training time of algorithms using the same ranking models, as described next. 

\begin{itemize}

    \item  \textit{Pairwise and enriched datasets:}  DQN $>$  PPO2 $>$  ACKTR $=$ A2C $>$ ACER $>$ TRPO $=$ PPO1
    \item  \textit{Pointwise and enriched datasets:}  SAC $>$ TD3 $=$ A2C $<$ DDPG $>$ PPO1 $>$  ACKTR $>$ PPO2 $>$  TRPO 
    \item  \textit{Listwise and enriched datasets:}  DQN $>$ PPO1 $>$ TRPO $>$ PPO2 = A2C =  ACER = DQN 
    \item  \textit{Pairwise and simple datasets:}  DQN $>$ ACKTR $>$ A2C = PPO1 = PPO2 = ACER  $>$ TRPO 
    \item  \textit{Pointwise and simple datasets:} SAC =  DDPG  = TD3 $>$ A2C = PPO1 = PPO2 =  ACKTR = TRPO 
    \item  \textit{Listwise and simple datasets:} DQN $>$ PPO1 $>$ TRPO $>$  ACER = PPO2 = A2C 

\end{itemize}

\TR{E3, R2.2, R2.6, R3.6}{
To compare prediction times, we perform an analysis similar to the one above at each cycle. Results clearly show that there is a significant difference across the prediction time of algorithms using the same ranking models, as described next.}

{\color{black}
\begin{itemize}

    \item  \textit{Pairwise and enriched datasets:}  A2C $=$ PPO1 $<$ PPO2 $<$ DQN $=$ ACKTR $<$  TRPO $<$ ACER

    \item  \textit{Pointwise and enriched datasets:} A2C $<$ DDPG $=$ PPO1 $<$ PPO2 $<$ TD3 $<$ ACKTR $<$ SAC $<$ TRPO

    \item  \textit{Listwise and enriched datasets:}   DQN $<$ A2C $<$ ACER $<$ PPO1 $<$ PPO2 $<$ TRPO

    \item  \textit{Pairwise and simple datasets:} A2C $<$ PPO1  $<$ PPO2  $<$ DQN  $=$ 'ACKTR'  $<$ TRPO  $<$ ACER

    \item  \textit{Pointwise and simple datasets:} PPO1 $>$ DDPG $>$ PPO2 $>$ A2C: $=$ TD3 $>$ ACKTR $=$ SAC $>$ TRPO: 6

    \item  \textit{Listwise and simple datasets:} A2C $=$ ACER $<$ TRPO $<$ PPO2 $<$ PPO1

\end{itemize}
}

\TR{E3, R2.2, R2.6, R3.6}{
Based on the above results, we can conclude that: 1) DQN, SAC, and TD3 are the worst algorithms in terms of training time, and 2) listwise configurations feature the worst prediction times. Since the accuracy of listwise configurations, as mentioned earlier, is relatively low regardless of the algorithm, we can therefore recommend against their use for test case prioritization. In contrast, overall PPO2, A2C, and ACER are relatively fast to train, as their prediction time is less than 2.22 seconds for pairwise and pointwise. Further, since their accuracy, especially that of ACER, is relatively good using pairwise ranking, we can recommend using them with the pairwise model for test case prioritization.  
}

\TR{R3.5}{Further, from the results of RQ1.3, we can see that one of the actor-critic algorithms always provides the best result in terms of accuracy, training time, and prediction time.  
Actor-critic algorithms combine the strong points of actor-only and critic-only methods. The critic estimates the value function, and the actor updates the policy distribution in the direction suggested by the critic (such as with policy gradients)~\cite{konda2000actor}. Also, these algorithms, such as ACER and PPO2, use multiple workers to avoid the use of a replay buffer and this results in a faster training time~\cite{A2C}.}

\subsubsection{RQ2}


\begin{table}[t]
\caption{Welch's t-test Results and Common Language Effect Size between Pairwise-ACER and Baselines.}
\label{tab:baselinecompare}
\begin{center}
\begin{tabular}{ p{1.1cm}|cccccc} 
\hline
\multicolumn{1}{c|}{\textbf{}}  & 
\multicolumn{2}{c}{\small \textbf{RL-BS1.}} & 
\multicolumn{2}{c}{\small \textbf{RL-BS2} }& 
\multicolumn{2}{c}{\small \textbf{MART}}  
 
\\
&  
\small \textit{p}-val & \small CLE &
\small \textit{p}-val & \small CLE &

\small \textit{p}-val & \small CLE
\\
\hline
\hline

\small \textbf{IO} & \small  NA  & \small  NA &   \small .0000 & \small .985 & \small .0000 &  \small .931 \\
\small \textbf{CODEC} & \small  NA  & \small  NA &  \small .0000 & \small .942  & \small  \small .1057 &  \small .551 \\
\small \textbf{IMAG} & \small  NA  & \small  NA & \small .0000 & \small .895  &\small .0000 &  \small .854  \\
\small \textbf{COMP} & \small  NA  & \small  NA  & \small .0000 & \small .976   & \small .0000 &  \small .795  \\
\small \textbf{LANG} & \small  NA  & \small  NA &  .0000 & \small .931  & \small .0000 &  \small .677  \\
\small \textbf{MATH} & \small  NA  & \small  NA &  .0000 & \small .915 & \small .1218 &  \small .681 \\
 
\small \textbf{Paint.} & \small .7483 & \small .486 &  \small  NA  & \small  NA & \small  NA & \small NA \\

\small \textbf{IOFROL} & \small .0000 & \small .376 &  \small  NA  & \small  NA  & \small  NA & \small NA \\

\hline
\hline

\end{tabular}
\end{center}

\end{table}
\textbf{Overview.} The averages and standard deviations of baselines for NRPA and APFD are shown in the last three rows of Table~\ref{tab:results}, for the datasets on which they were originally experimented. We used one of the best configurations (highest average accuracy) from RQ1 across all datasets (pairwise-ACER) and compared the results of this configuration with the baselines in terms of NRPA or APFD.  As before, we used the Welch t-test since it does not assume equal variances. We calculate  CLE again, this time between pairwise-ACER and baselines to assess the effect size of differences. Table~\ref{tab:baselinecompare} shows the results of Welch t-test and CLE for all datasets. 

\textbf{RQ2.1}
 Row RL-BS1 of Table~\ref{tab:results} corresponds to the results of the RL-based solution proposed by Spieker et al.~\cite{spieker2017reinforcement}. We have replicated their experiment and calculated APFD, since no detailed results were available online. As discussed in Section ~\ref{sec:baselines}, this work originally relies on simple history data and, therefore, we did not attempt to apply it to enriched datasets. As shown in the first two columns of row RL-BS1,  for dataset Paint-Control, pairwise-ACER performs as well as RL-BS1, i.e., there are no statistically significant differences in results. Also, for dataset IOFROL, RL-BS1 fares slightly better than pairwise-ACER, however with a low CLE of 62.4. But overall, no approach (pairwise-ACER or RL-BS1) performs well, more particularly with IOFROL. It is interesting to note that the average APFD of RL-BS1 and pairwise-ACER (1) are around 0.15 lower than that of the optimal ranking (row \textit{Optimal} of Table~\ref{tab:results}), and (2) are not significantly different from a simple heuristic solution that would prioritize test cases based on their recent verdicts with recently failed test cases assigned a higher priority (APFDs 0.632 and 0.772, for IOFROL and Paint-Control, respectively). These results suggest that simply relying on test execution history, such as the one available in simple datasets, does not provide sufficient features for learning an accurate test prioritization policy.  
 
 
Row RL-BS2 shows the results of the best configuration (RL-BS2) among the RL-based solutions proposed by Bertolino et al. ~\cite{bertolinolearning}. We have used the detailed experimental results available online to recalculate NRPA while ignoring the cycles with less than five test cases. As shown in the corresponding row,  pairwise-ACER fares significantly better for all datasets. To check that the differences in average are statistically significant, we performed again a Welch t-test for each dataset. As shown in Table~\ref{tab:baselinecompare}, pairwise-ACER is significantly better for all datasets.  Further, CLE ranges between 0.89 to 0.98, which implies that for at least 89\% of the cycles, pairwise-ACER performs better than RL-BS2. 
  
Thus, according to the results above, we can safely conclude that pairwise-ACER significantly improves, in terms of ranking accuracy, the state of the art regarding the use of RL for test case prioritization. 
Since the baselines use only one episode of training per cycle, their training time is significantly lower than our best configuration (pairwise-ACER), which is based on the pairwise ranking model. But as shown in Figure~\ref{fig:trainingtimescer}, the average execution time per cycle is less than 5 minutes  across all datasets and the worst-case training time is less 25 minutes. Since our approach uses offline training by replaying logs and enables the training process to be run in the background, such training times would not add any delay to the CI process. Therefore, the extra training time of  pairwise-ACER compared to baselines, which is in the order of minutes, does not have practical consequences. 

\begin{figure}
    \centering
    \includegraphics[width=8cm]{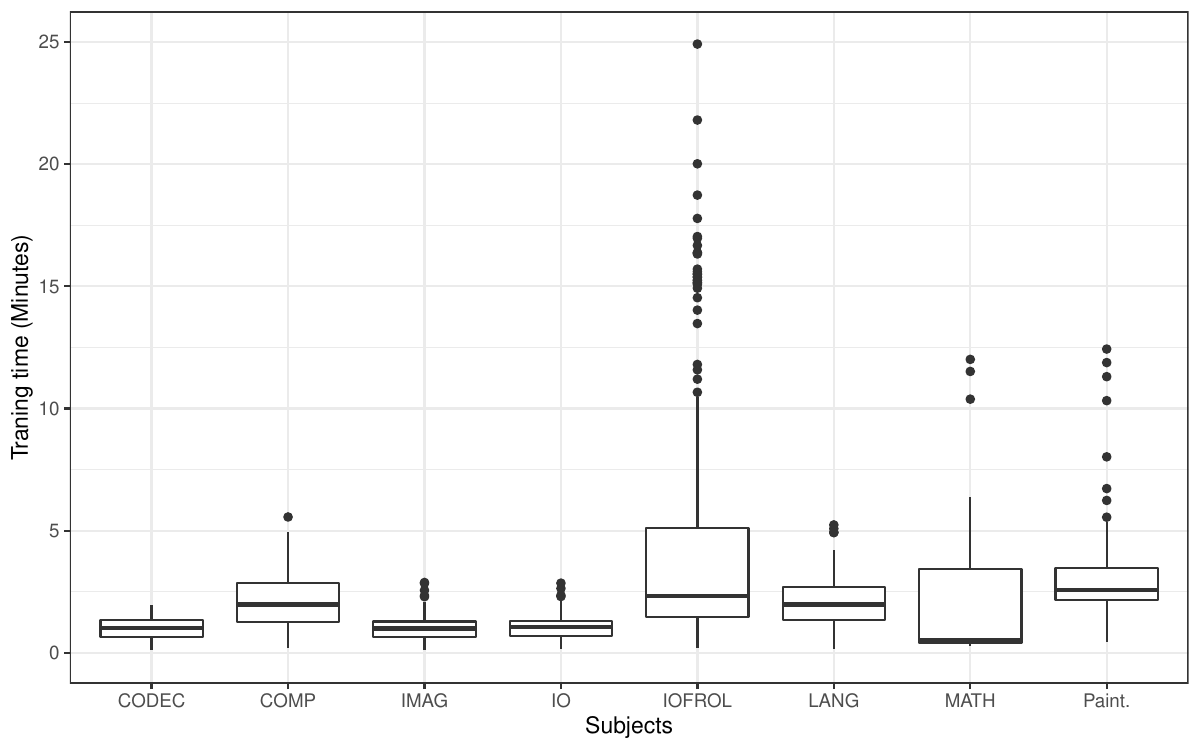}
    \caption{Training time of pairwise-ACER for all datasets}
    \label{fig:trainingtimescer}
\end{figure}

\textbf{RQ2.2}
Row MART (MART ranking model) in Table~\ref{tab:results} provides the results of the best ML-based solution according to a comprehensive evaluation~\cite{bertolinolearning}. For all datasets, except CODEC and MATH where results are equivalent, pairwise-ACER performs better than MART. We once again performed a Welch t-test and, as shown in Table~\ref{tab:baselinecompare}, the test results confirm that the differences for CODEC and MATH are not significant, while they are significant for the other four datasets. Also, to measure the effect size, we calculated CLE, as shown in Table~\ref{tab:baselinecompare}. The CLE of pairwise-ACER vs. MART ranges between 0.551 to 0.931 with an average of 0.75, i.e., in 75\% of the cycles, pairwise-ACER fares better than MART. Therefore, we can safely conclude that pairwise-ACER advances the state of art 
compared to the best ML-based ranking technique (MART). 
 

In addition to their higher ranking accuracy over MART, RL agents can be incrementally trained to adapt to changes in the system and test suites, which is one of the main challenges in the context of frequently-changing and dynamic CI environments, in which new execution logs and code changes are continuously collected. In contrast, the MART ranking model is an ensemble model of boosted regression trees. Boosting algorithms, as a class of ensemble learning methods, are designed for static training, based on a fixed training set. Thus, they cannot be directly and easily applied to online learning and incremental learning~\cite{zhang2019incremental}. Supporting incremental learning in boosting algorithms is an active research area for which no solution is currently available in existing libraries~\cite{Ranklibs}. This causes a practical issues since the performance of the ranking model will gradually decay after some cycles, and a new model needs to be trained based on the most recent data. In contrast, RL algorithms gradually adapt based on the incoming data and there is no need to train a new model from scratch. 

 As mentioned above, previous attempts to apply RL to test case prioritization had brought adaptability at the expense of accuracy. In this work, pairwise ranking using the ACER algorithm has significantly improved ranking accuracy over that of MART, the best reported ML-based ranking model for test case prioritization, as well as over that of previous attempts at using RL. Thus, we can conclude that best RL configurations, for example based on a pairwise ranking model and the ACER algorithm, have the potential to be a reliable and adaptive solution for test case prioritization in CI environments.
}

\subsection{Threats to validity} 

\TR{E2, R1.4, R2.5, R2.1} {The low and high failure rates of enriched and simple data sets, respectively, may threaten the generality of our evaluation results. However, we only use these data sets to make a fair comparison with related work under identical circumstances. We do not make any general claim on the effectiveness of our approach in absolute terms but rather focus on relative effectiveness across ranking models and RL techniques. Note that the fact that we get largely consistent results (in terms of trends, not magnitudes) with low and high failure rates is reassuring regarding the generalizability of the recommendations we provide regarding ranking models and RL algorithms. 
}


{\color{black}
Further, low failure rates such as the one for the enriched dataset tend to characterize the datasets typically used for evaluating test selection and prioritization techniques~\cite{do2006use,luo2018assessing}. Further, in a CI context, Beller et al.~\cite{beller2017oops} conducted a comprehensive analysis of TravisCI projects and showed that for all 1,108 Java projects with test executions, the ratio of builds with at least one failed test case has a median of 2.9\% and a mean of 10.3\%. To deal with this issue, for evaluation purposes, studies focused on non-ML techniques rely on seeded faults, which are typically produced through hand-seeding or program mutation fault injection techniques~\cite{do2006use,luo2018assessing}. In the context of ML-based techniques, where the goal is to train an ML model based on the history of test executions and source code changes, using fault injection techniques is not a valid option since it would add some faults randomly into the system, that have no relation with history. 

Another potential threat to validity is related to our evaluation metrics, which are standard across existing studies. We, however, in Section~\ref{sec:metrics}, discuss their limitations, how they should be interpreted, and when they should be used. 

}


\section{Conclusion}
\label{sec:conclusion}

In this paper, we formalized and investigated test case prioritization in continuous integration (CI) environments as a Reinforcement Learning (RL) problem. Our main motivation is to benefit from the capacity of RL to seamlessly adapt to changes in systems and test suites, while potentially reaching high ranking accuracy of regression test cases. Such high accuracy would help detect as many regression faults as quickly as possible within the tight available resources that are typically available in CI contexts, where frequent changes take place. 

Formalization is guided by the three well-known ranking models from the information retrieval domain: pairwise, pointwise, and listwise. Further, we have implemented this formalization by using a diverse set of carefully selected, state-of-the-art RL algorithms. 

We then performed an extensive evaluation over eight subject systems by combining 10 RL algorithms with the three ranking models, that resulted in 21 RL configurations. The evaluation reveals that by using a pairwise ranking model and the ACER algorithm~\cite{ACER}, an actor critic-based RL algorithm (pairwise-ACER), we obtain the best ranking accuracy. This accuracy, when enriching test execution history data with light-weight code features, is furthermore very close to the optimal ranking of test cases based on the actual failure data and execution times.

To position our work, we have compared pairwise-ACER with the two recent RL approaches and the best ML solution (MART) reported for test case prioritization. Using the standard NRPA ranking accuracy metric (ranging from 0 to 1), based on enriched datasets, the results show a significant ranking improvement when compared with both previous RL-based work (+0.1 on average) and MART (+0.027 on average). Further, we reach very high accuracy (NRPA $>$ 0.96), thus enabling the application of RL in practice. Simple datasets only based on execution history do not lead, with any learning technique, to satisfactory results. Differences in training time across approaches are not practically relevant. Based on our results, we conclude that the use of our optimal RL configuration (pairwise-ACER) can provide, based on adequate history and code data, a reliable and adaptive solution for test case prioritization in CI environments. 

While our work advances the state of the art in the use of RL techniques for test case prioritization, a certain number of issues remains open that should be tackled by future work.  In the following, we discuss the most important ones.
\begin{itemize}
    \item Tuning and optimization of our current approach. As discussed earlier, RL algorithms come with a set of hyperparameters that need to be tuned. However, we applied all of the algorithms with their default hyperparameters. Also, we only evaluated a limited number of reward functions that impact the performance of RL algorithms. Thus, optimization and tuning the best configuration of our approach (pairwise-ACER) is a natural next step to this work that can be performed automatically and systematically using search-based tuning frameworks such as Optuna~\cite{optuna_2019}.    
    \item Preparing of a rich dataset and a benchmark. Though, for comparison purposes, we used existing datasets that were made available by  previous studies, we also observed that the available datasets are limited in terms of features, the number of products, and the diversity of failure rates (failure rates in our benchmarks are either very low or very high). Also, there is no available benchmark for comparing available techniques based on a set of identical, representative datasets. As an effort in this direction, we have been working on the analysis and extraction of detailed test case execution data and source code history, via extending TravisTorrent~\cite{msr17challenge}, a tool for analysis of build logs of systems using Travis CI.  
 
\end{itemize}

Further, as we discussed, the pairwise ranking of test cases using ACER algorithm provides better results compared to the state of the art ranking libraries in the context of test case prioritization. It would be interesting to perform extended evaluation between the two methods to check whether or not the similar results can be achievable in a more general context, i.e., pairwise ranking fares better than the state of the art ranking model in a more general context.

\section*{Acknowledgement}
This work was supported by a research grant from Huawei Technologies Canada Co., Ltd, as well as by the Canada Research Chair and Discovery Grant programs of the Natural Sciences and Engineering Research Council of Canada (NSERC).

We also want to express our gratitude to the authors of the two previous studies on reinforcement learning and test prioritization (~\cite{bertolinolearning,spieker2017reinforcement}), for making their data and artifacts available and answering our questions.

\bibliographystyle{IEEEtran}
\bibliography{main.bib}

\end{document}